\documentclass[11pt]{article}
\usepackage[english,activeacute]{babel}
\usepackage{natbib}
\usepackage{comment}
\usepackage{float}
\usepackage[hidelinks]{hyperref}
\usepackage{mathrsfs}
\usepackage{enumitem}
\usepackage[font={small,it}]{caption}
\usepackage{amsmath,amsfonts,amsthm,amssymb}
\usepackage{bm,rotating,multirow,dsfont,graphicx}
\usepackage[usenames, dvipsnames]{color}
\usepackage{url}
\usepackage{multicol}
\usepackage{multirow}
\usepackage[T1]{fontenc}
\usepackage{flafter}
\usepackage{appendix}
\usepackage{subfigure}
\usepackage{xcolor}

\hypersetup{
	colorlinks   = true, 
	urlcolor     = blue, 
	linkcolor    = blue, 
	citecolor    = black 
}
\hypersetup{
	colorlinks,
	linkcolor={black}, 
	citecolor={black},
	urlcolor ={black}
}

\makeatletter
\def\hlinewd#1{%
	\noalign{\ifnum0=`}\fi\hrule \@height #1 %
	\futurelet\reserved@a\@xhline}
\makeatother

\def\obs{\textsf{obs}}
\def\nobs{\textsf{nobs}}
\def\@roman#1{\romannumeral #1}

\begin{document}

\def\spacingset#1{\renewcommand{\baselinestretch}{#1}\small\normalsize}\spacingset{1}

\title{Small area estimation using multiple imputation in three-parameter logistic models}

\date{}
	
\author{
    Cristian Tellez-Piñerez, Universidad de Santo Tomás, Colombia\footnote{Email: cristiantellez@usantotomas.edu.co} \\
    Leonardo Trujillo,       Universidad Nacional de Colombia, Colombia \\
    Andrés Gutiérrez-Rojas,  CEPAL, Chile \\
	Juan Sosa,               Universidad Nacional de Colombia, Colombia
}	 
	       
\maketitle

\begin{abstract} 
We propose a novel methodology relating item response theory methods with small area estimation strategies in the presence of missing data. Specifically, we propose an unbiased estimator for the average ability parameter of three-parameter logistic models. Thus, we carry out an extensive simulation study in order to compare our estimator with the well-known Horvitz-Thompson estimator. According to our experiments with synthetic data, our proposal has substantial lower standard errors than its competitor. In addition, we perform an actual application by considering the Mathematics results of the 2015 Program for International Student Assessment (PISA), and also, compare our results with previous analyses. Our findings strongly suggest that our methodology is a high competitive alternative for generating compelling official statistics.
\end{abstract}


\spacingset{1.1} 

\section{Introduction}

Educational assessment can be understood as the process of using collected information about attitudes, beliefs, knowledge, and skills to improve learning in academic programs \citep{Allen}. 
These data are typically obtained from standardized tests applied to students for assessing planned learning goals (\citealt{Kuh}, \citealt{ManCal}, \citealt{PISA_2015}, \citealt{UNESCO}). 
However, standardized tests for educational assessment are currently facing a serious issue: A decreasing number of participants per application due to lack of access and financial restrictions.

A common problem is the presence of missing data in response strings. 
For example, in the Colombian institute for educational assessment, previous test applications as well as findings included in technical manuals (e.g., \citealt{ICFES7}), show that students would take more than fifteen hours to go through all the questions in the entire test. 
Exposing each student to a fifteen-hours test is not only a counterproductive strategy for the final scores, but also promotes cognitive conflicts  from a pedagogical point of view \citep{alvarez2007desarrollo}.
This is also the case with other tests, including Program for International Student Assessment (PISA, \citealt{Reporte_tecnico_PISA_2012}), Trends in the International Study of Mathematics and Sciences (TIMMS, \citealt{mullis2015assessment}), and Saber 3$^\text{o}$, 5$^\text{o}$ y 9$^\text{o}$ tests \citep{ManCal}, among many others.

Estimation and analysis of standardized tests are mostly carried out employing Item Response Theory (IRT) methods (e.g., \citealt{lord1980applications}, \citealt{martinez1995psicometria}, \citealt{fernandez1997introduccion}, \citealt{hambleton2013item}). 
IRT models intend to explain the relationship between latent traits, i.e., unobservable characteristics or attributes, and their manifestations, i.e., observed outcomes, responses or performance \citep{embretson2013item}.
In particular, three-parameters logistic models (3PLMs, e.g., \citealt{paek2019using}) are very popular for such and end.
According to \citet{sulis}, these models characterize the probability of responses to any particular question (item) as a function of a location parameter (basal position along the latent trait), a discrimination parameter (item's capability to discriminate individuals with different latent trait values), and finally, an ability parameter (intensity of the latent trait).

To the best of our knowledge, there is a substantial lack of research about non-response methods in the context of standardized tests (e.g., \citealt{adams}, \citealt{baker}, \citealt{sulis}). 
Therefore, our foundational aim relies on extending IRT methodologies accounting for missing data, while achieving similar quality indicators at lower costs.
Such a task is displayed in Table \ref{tabla_proporcion-1-2}. 
Rows represent students who are probabilistically chosen to take a given test. 
Students are grouped by domains, which may represent educational establishments. 
Columns are divided in two classes, namely, strings and plausible values. 
String columns register the response of each student as 1 if a given question was answered correctly, as 0 if not, and finally as ``x'' for all those questions not applied to a particular student.
Non-applied questions are drawn according to specific random schemes 
(e.g., the \textit{Instituto Colombiano para la Evaluaci\'on de la Educaci\'on}, ICFES, uses balanced incomplete blocks for that end, \citealt{en2017Armado}). 
Plausible values columns contain the realization of random variables that estimate each student's ability to answer a given question accounting for the presence of missing data.
Thus, there are available $k$ plausible values for each student and $n_d$ random variable realizations for each plausible value, which implies that $n_d \times k$ values are available in domain $d$, for $d = 1,\ldots,D$.
Finally, within each domain $d$, we have $k$ estimates of the ability mean along with $k$ estimates of the corresponding variance.

\begin{table}[!ht]
\begin{centering}
{\tiny{}}%
\begin{tabular}{ccccccccccc}
\hline 
\textbf{\tiny{}Domain} & \textbf{\tiny{}Student} & \multicolumn{5}{c}{\textbf{\tiny{}String}} & \multicolumn{4}{c}{\textbf{\tiny{}Plausible values}}\tabularnewline
\hline 
\multirow{3}{*}{{\tiny{}1}} & {\tiny{}1} & {\tiny{}x} & {\tiny{}0} & {\tiny{}x} & {\tiny{}$\cdots$} & {\tiny{}1} & {\tiny{}$\tilde{\theta}_{111}$} & {\tiny{}$\tilde{\theta}_{112}$} & {\tiny{}$\cdots$} & {\tiny{}$\tilde{\theta}_{11k}$}\tabularnewline
 & {\tiny{}$\vdots$} & \multicolumn{5}{c}{{\tiny{}$\vdots$}} & \multicolumn{4}{c}{{\tiny{}$\vdots$}}\tabularnewline
 & {\tiny{}$n_{1}$} & {\tiny{}1} & {\tiny{}x} & {\tiny{}0} & {\tiny{}$\cdots$} & {\tiny{}x} & {\tiny{}$\tilde{\theta}_{1n_{1}1}$} & {\tiny{}$\tilde{\theta}_{1n_{1}2}$} & {\tiny{}$\cdots$} & {\tiny{}$\tilde{\theta}_{1n_{1}k}$}\tabularnewline
 &  &  &  &  &  &  &  &  &  & \tabularnewline
{\tiny{}Direct estimation} &  &  &  &  &  &  & {\tiny{}$\hat{\theta}_{11}^{\textsf{dir}}$} & {\tiny{}$\hat{\theta}_{12}^{\textsf{dir}}$} & {\tiny{}$\cdots$} & {\tiny{}$\hat{\theta}_{1k}^{\textsf{dir}}$}\tabularnewline
{\tiny{}Estimated variance} &  &  &  &  &  &  & {\tiny{}$\widehat{\textsf{Var}}\left(\hat{\theta}_{11}^{\textsf{dir}}\right)$} & {\tiny{}$\widehat{\textsf{Var}}\left(\hat{\theta}_{12}^{\textsf{dir}}\right)$} & {\tiny{}$\cdots$} & {\tiny{}$\widehat{\textsf{Var}}\left(\hat{\theta}_{1k}^{\textsf{dir}}\right)$}\tabularnewline
\hline 
\multicolumn{11}{c}{{\tiny{}$\vdots$}}\tabularnewline
\hline 
\multirow{3}{*}{{\tiny{}$d$}} & {\tiny{}1} & {\tiny{}x} & {\tiny{}1} & {\tiny{}1} & {\tiny{}$\cdots$} & {\tiny{}x} & {\tiny{}$\tilde{\theta}_{d11}$} & {\tiny{}$\tilde{\theta}_{d12}$} & {\tiny{}$\cdots$} & {\tiny{}$\tilde{\theta}_{d1k}$}\tabularnewline
 & {\tiny{}$\vdots$} & {\tiny{}$\vdots$} &  &  &  &  &  &  &  & \tabularnewline
 & {\tiny{}$n_{d}$} & {\tiny{}0} & {\tiny{}x} & {\tiny{}0} & {\tiny{}$\cdots$} & {\tiny{}1} & {\tiny{}$\tilde{\theta}_{dn_{d}1}$} & {\tiny{}$\tilde{\theta}_{dn_{d}2}$} & {\tiny{}$\cdots$} & {\tiny{}$\tilde{\theta}_{dn_{d}k}$}\tabularnewline
 &  &  &  &  &  &  &  &  &  & \tabularnewline
{\tiny{}Direct estimation} &  &  &  &  &  &  & {\tiny{}$\hat{\theta}_{d1}^{\textsf{dir}}$} & {\tiny{}$\hat{\theta}_{d2}^{\textsf{dir}}$} & {\tiny{}$\cdots$} & {\tiny{}$\hat{\theta}_{dk}^{\textsf{dir}}$}\tabularnewline
{\tiny{}Estimated variance} &  &  &  &  &  &  & {\tiny{}$\widehat{\textsf{Var}}\left(\hat{\theta}_{d1}^{\textsf{dir}}\right)$} & {\tiny{}$\widehat{\textsf{Var}}\left(\hat{\theta}_{d2}^{\textsf{dir}}\right)$} & {\tiny{}$\cdots$} & {\tiny{}$\widehat{\textsf{Var}}\left(\hat{\theta}_{dk}^{\textsf{dir}}\right)$}\tabularnewline
\hline 
\multicolumn{11}{c}{{\tiny{}$\vdots$}}\tabularnewline
\hline 
\multirow{3}{*}{{\tiny{}$D$}} & {\tiny{}1} & {\tiny{}1} & {\tiny{}0} & {\tiny{}x} & {\tiny{}$\cdots$} & {\tiny{}1} & {\tiny{}$\tilde{\theta}_{D11}$} & {\tiny{}$\tilde{\theta}_{D12}$} & {\tiny{}$\cdots$} & {\tiny{}$\tilde{\theta}_{D1k}$}\tabularnewline
 & {\tiny{}$\vdots$} &  &  &  &  &  &  &  &  & \tabularnewline
 & {\tiny{}$n_{D}$} & {\tiny{}0} & {\tiny{}x} & {\tiny{}0} & {\tiny{}$\cdots$} & {\tiny{}1} & {\tiny{}$\tilde{\theta}_{Dn_{D}1}$} & {\tiny{}$\tilde{\theta}_{Dn_{D}2}$} & {\tiny{}$\cdots$} & {\tiny{}$\tilde{\theta}_{Dn_{D}k}$}\tabularnewline
 &  &  &  &  &  &  &  &  &  & \tabularnewline
{\tiny{}Direct estimation} &  &  &  &  &  &  & {\tiny{}$\hat{\theta}_{D1}^{\textsf{dir}}$} & {\tiny{}$\hat{\theta}_{D2}^{\textsf{dir}}$} & {\tiny{}$\cdots$} & {\tiny{}$\hat{\theta}_{Dk}^{\textsf{dir}}$}\tabularnewline
{\tiny{}Estimated variance} &  &  &  &  &  &  & {\tiny{}$\widehat{\textsf{Var}}\left(\hat{\theta}_{D1}^{\textsf{dir}}\right)$} & {\tiny{}$\widehat{\textsf{Var}}\left(\hat{\theta}_{D2}^{\textsf{dir}}\right)$} & {\tiny{}$\cdots$} & {\tiny{}$\widehat{\textsf{Var}}\left(\hat{\theta}_{Dk}^{\textsf{dir}}\right)$}\tabularnewline
\hline 
\end{tabular}{\tiny\par}
\par\end{centering}
\caption{Information per domain and per student. {\tiny{}\label{tabla_proporcion-1-2}}}
\end{table}

In contrast with standard methods, our proposal incorporates estimates of the ability parameter using the Fay-Herriot approach \citep{fay1979estimates} within the framework of IRT modeling where multiple imputation tasks are needed. 
Such a methodology has in itself two major contributions. On the one hand, it has profound practical implications to deal properly with problematic data structures involving missing data, and on the other, it resolves challenging theoretical issues associated with the production of reliable official statistics. 
We highlight that this is a complex task demanding auxiliary variables correlated with the ability of the students (such as parental socio-economic level, parental education, school infrastructure, among others, according to \citealt{trevino2010factores}), as well as sophisticated statistical tools for computing the resulting estimator together with its Mean Square Error (MSE).

This article is structured as follows: Section \ref{sec_theoretical_development} shows the theoretical
development of our estimator along with its MSE under the restrictions presented above; 
Section \ref{sec_simulation_study} presents an exhaustive simulation study in which the proposed estimator
is compared with existing estimators in the literature; Section \ref{sec_application} exhibits an application of our methodology regarding the 2015 PISA test. Finally, Section \ref{sec_discussion} discusses our findings as well as some relevant aspects for future research.

\section{Theoretical development of the estimator}\label{sec_theoretical_development}

In this section, we present our approach for estimating the average in small areas using multiple imputation in three-parameter logistic models (3PLMs, e.g., \citealt{paek2019using}).
Firstly, we review the estimation of plausible values. Secondly, we use small area estimation theory as well as the Fay-Herriot model in order to obtain expressions of the ability mean estimator along with its Mean Square Error (MSE).

\subsection{Ability estimation using plausible values and 3PLMs}

We have a finite universe $U = \bigcup_{d=1}^{D}U_{d}$ of size $N=\sum_{d=1}^{D}N_{d}$ distributed in $d$ domains, where $U_d$ is the population in domain $d$ of size $N_d$. 
Also, let $s = \bigcup_{d=1}^{D}s_{d}$ be the sample of size $n=\sum_{d=1}^{D}n_{d}$ under consideration, where $s_d$ is the sample in domain $d$ of size $n_{d}$ drawn under a particular sampling design $p(\cdot)$. 
It may happen that for some domain $d$, either $s_{d}$ turns out to be empty or $n_d$ is not large enough in order to enable reliable estimations. 
Moreover, indicator variables $\xi_{ij}$ registering whether individual $j$ answers item $i$ correctly, $\xi_{ij} = 1$, or not, $\xi_{ij} = 0$, are observed, for $j=1,\ldots,n_{d}$ and $i=1,\ldots,I$.

IRT modeling assumes that individuals are endowed with an ability to answer items correctly. Let $\boldsymbol{\theta}$ be an array containing the ability parameters of all the individuals under consideration.
In order to find the distribution of $\boldsymbol{\theta}$, there must be available known individual-level auxiliary information, typically stored in a vector of variables $\boldsymbol{x}_{j}^{\textsf{I}}$ for each $j=1,\ldots,n_{d}$ (the sub-index $\textsf{I}$ emphasizes the notion of auxiliary information at the level of individuals or subjects).
Thus, the probability distribution of $\boldsymbol{\theta}$ in the population is not only conditional on the observed indicator variables $\boldsymbol{\xi}_{\obs}$, but also on the auxiliary information $\mathbf{X}_{\textsf{I}}$, i.e.,
\begin{equation*}
\textsf{p}\left(\boldsymbol{\theta}\mid\boldsymbol{\xi}_{\obs},\,\mathbf{X}_{\textsf{I}}\right) \propto 
\textsf{p}\left(\boldsymbol{\xi}_{\obs}\mid \boldsymbol{\theta},\mathbf{X}_{\textsf{I}}\right)
\textsf{p}\left(\boldsymbol{\theta}\mid\mathbf{X}_{\textsf{I}}\right)\,,    
\end{equation*}  
where $\mathbf{X}_{\textsf{I}}$ is a matrix storing all the individual-level auxiliary information.
Assuming conditional independence between $\boldsymbol{\xi}_{\obs}$ and $\mathbf{X}_{\textsf{I}}$ as in \citet{rubin1991multiple}, which is quite reasonable in practice, it is easy to see that 
\begin{equation*}
\textsf{p}(\boldsymbol{\theta}\mid\boldsymbol{\xi}_{\obs},\mathbf{X}_{\textsf{I}}) \propto \textsf{p}\left(\boldsymbol{\xi}_{\obs}\mid\boldsymbol{\theta}\right)\,
\textsf{p}\left(\boldsymbol{\theta}\mid\mathbf{X}_{\textsf{I}}\right)\,.
\end{equation*}

Under this setting, the main goal relies on finding the conditional distribution of $\boldsymbol{\theta}$ given $\boldsymbol{\xi}_{\obs}$ and $\mathbf{X}_{\textsf{I}}$, which in turn
depends on two conditional distributions.
Firstly, $\textsf{P}(\boldsymbol{\xi}_{\obs}\mid\boldsymbol{\theta})$, 
the response chain distribution of students given their ability, considered here as a standard 3PLM \citep{bock1981marginal}.
Secondly, $\textsf{P}(\boldsymbol{\theta}\mid\mathbf{X}_{\textsf{I}})$, the ability distribution of students given the auxiliary information, regarded here as a multivariate Normal distribution with mean $\mathbf{X}_{\textsf{I}}\boldsymbol{\Gamma}$ and covariance matrix $\mathbf{\Sigma}$, where both $\boldsymbol{\Gamma}$ and $\mathbf{\Sigma}$ need to be estimated. On this point, we are implicitly stating that the parameter space of each component of $\boldsymbol{\theta}$ is the real line, even though most of the ability mass lies between $-3$ and $3$ (e.g., \citealt{de2000teoria}).

It is straightforward to see that the conditional distribution $\textsf{p}(\boldsymbol{\theta}\mid\boldsymbol{\xi}_{\obs},\mathbf{X}_{\textsf{I}})$ is completely determined by handling the unknown parameters in $\textsf{p}(\boldsymbol{\xi}_{\obs}\mid\boldsymbol{\theta})$ as well as those in $\textsf{p}(\boldsymbol{\theta}\mid\mathbf{X}_{\textsf{I}})$.
Existing IRT methods typically deal with this setting by, 
first, using the Expectation–Maximization (EM) algorithm (e.g., \citealt{bock1981marginal}) to estimate the unknown parameters in $\textsf{P}(\boldsymbol{\theta}\mid\mathbf{X}_{\textsf{I}})$, and
then, using the Metropolis-Hastings (MH) algorithm (e.g., \citealt{fox2010bayesian}) to draw $L$ plausible values (i.e., abilities estimates) for each individual $j$ in domain $d$. 
In this spirit, let $\theta_{dj\ell}^{\textsf{pv}}$ be the $\ell$-th MH plausible value (hence the upper-index $\textsf{pv}$) of individual $j$ in domain $d$, for $\ell=1,\ldots,L$, with $j=1,\ldots,J$ and $d=1,\ldots,D$. 
Also, let $\gamma_{d}=g(\theta_{dj\ell}^{\textsf{pv}})$ be an arbitrary function of $\theta_{dj\ell}^{\textsf{pv}}$ characterizing a specific feature in the $d$-th domain (e.g., the mean). Thus, an estimate of $\gamma_d$ and its corresponding variance can be found by noticing that
\begin{equation*}
\textsf{p}(\gamma_{d}\mid\boldsymbol{\xi}_{\obs}) = 
\int \textsf{p}\left(\gamma_d\mid\boldsymbol{\xi}_{\nobs},\boldsymbol{\xi}_{\obs}\right)\,
\textsf{p}\left(\boldsymbol{\xi}_{\nobs}\mid\boldsymbol{\xi}_{\obs}\right)\,\textsf{d}\boldsymbol{\xi}_{\nobs}\,,
\end{equation*}
where $\boldsymbol{\xi}_{\nobs}$ is  composed of all those unobserved indicators variables not given in $\boldsymbol{\xi}_{\obs}$, and the integral is carried out over the space parameter of $\boldsymbol{\xi}_{\nobs}$. As a consequence, for the average of the individuals' abilities $\gamma_d$, taking $g(\cdot)$ as the mean of the $\theta_{dj\ell}^{\textsf{pv}}$, it follows that
\begin{eqnarray}
\textsf{E}\left(\gamma_{d}\mid\boldsymbol{\xi}_{\obs}\right) & = & \textsf{E}\left[\textsf{E}\left(\gamma_{d}\mid\boldsymbol{\xi}_{\nobs},\boldsymbol{\xi}_{\obs}\right)\mid\boldsymbol{\xi}_{\obs}\right]
\simeq \frac{1}{L}\sum_{\ell=1}^L\hat\theta_{d\ell}^{\textsf{pv}} = \hat{\gamma}_{d} \nonumber \\
\textsf{Var}\left(\gamma_{d}\mid\boldsymbol{\xi}_{\obs}\right) & = & \textsf{E}\left[\textsf{Var}\left(\gamma_{d}\mid\boldsymbol{\xi}_{\nobs},\boldsymbol{\xi}_{\obs}\right)\mid\boldsymbol{\xi}_{\obs}\right]+\textsf{Var}\left[\textsf{E}\left(\gamma_{d}\mid\boldsymbol{\xi}_{\nobs},\boldsymbol{\xi}_{\obs}\right)\mid\boldsymbol{\xi}_{\obs}\right]\nonumber\\
& \simeq &  \frac{1}{L} \sum_{\ell=1}^{L}\textsf{Var}(\hat\theta_{d\ell}^{\textsf{pv}}) + \left(1+\frac{1}{L}\right)\frac{1}{L-1}\sum_{\ell=1}^{L}(\hat\theta_{d\ell}^{\textsf{pv}}-\hat{\gamma}_{d})^{2}\,,
\label{Estimadores_directos_y_varianzas}
\end{eqnarray}
where 
$\hat\theta_{d\ell}^{\textsf{pv}}$ is the $\ell$-th plausible value in domain $d$, and
$\textsf{Var}(\hat\theta_{d\ell}^{\textsf{pv}})$ depends on the specific formulation of the sampling design and corresponds to the average of the estimated variances for each plausible value.

\subsection{Proposed estimator}

Here, we take a step further and go beyond the exiting IRT literature as presented above in order to carry out our main task:
Combining IRT methods with small area estimation strategies accounting for missing data. 
Thus, inspired in Small Area Estimation (SAE) methods, we propose estimating $\boldsymbol\gamma=(\gamma_1,\ldots,\gamma_D)$, as a result of the model
\begin{eqnarray}\label{eq_LMM}
\hat{\boldsymbol\gamma} & = & \mathbf{X}_{\textsf{A}}\boldsymbol{\beta}+\mathbf{Z}\boldsymbol{u}+\boldsymbol{e}\,,
\end{eqnarray}
where
$\mathbf{X}_{\textsf{A}}$ is a fixed-effects matrix storing all the area-level auxiliary information (the sub-index $\textsf{A}$ emphasizes the notion of auxiliary information at the level of areas or domains), 
$\boldsymbol{\beta}$ is a vector of unknown constants, 
$\mathbf{Z}$ is a random-effects design matrix, 
and finally, $\boldsymbol{u}$ and $\boldsymbol{e}$ are independent random vectors such that $\boldsymbol{u}\sim \textsf{N}\left(\boldsymbol{0},\,\mathbf{V}_{u}\right)$ and $\boldsymbol{e}\sim \textsf{N}\left(\boldsymbol{0},\,\mathbf{V}_{e}\right)$, with known $\mathbf{V}_{u}$ and $\mathbf{V}_{e}$.
Thus, under the previous specification it follows directly that $\textsf{Var}\left(\hat{\boldsymbol\gamma}\right)=\mathbf{Z}\mathbf{V}_{u}\mathbf{Z}^{\textsf{T}}+\mathbf{V}_{e}=\mathbf{V}$, 
and also, following standard results about linear mixed-effects models (e.g., \citealt{mcculloch2004generalized}), it can be shown that the optimal unbiased linear predictor of $\boldsymbol\tau=\mathbf{L}\boldsymbol{\beta}+\mathbf{M}\boldsymbol{u}$ is $\hat{\boldsymbol\tau}=\mathbf{L}\hat{\boldsymbol{\beta}}+\mathbf{M}\hat{\boldsymbol{u}},$ with 
$$
\hat{\boldsymbol{\beta}}=\left(\mathbf{X}_{\textsf{A}}^{\textsf{T}}\mathbf{V}^{-1}\mathbf{X}_{\textsf{A}}\right)^{-1}\mathbf{X}_{\textsf{A}}^{\textsf{T}}\mathbf{V}^{-1}\hat{\boldsymbol\gamma}
\qquad\text{and}\qquad
\hat{\boldsymbol{u}}=\left(\mathbf{Z}\mathbf{V}_{u}\right)^{\textsf{T}}\mathbf{V}^{-1}\left(\hat{\boldsymbol\gamma}-\mathbf{X}_{\textsf{A}}\hat{\boldsymbol{\beta}}\right)\,.
$$

Now, building on ideas given in \citet{harville1977maximum} and \citet{DomingoSAE}, we consider an unbiased linear predictor of the form $\hat{\boldsymbol\tau}=\mathbf{Q}_{0}+\mathbf{Q}_{1}\hat{\boldsymbol\gamma}$, where $\mathbf{Q}_{0}$ and $\mathbf{Q}_{1}$ are two conformable matrices. 
Since $\hat{\boldsymbol\tau}$ is unbiased, we have that $\textsf{E}\left(\hat{\boldsymbol\tau}-\boldsymbol\tau\right)=\mathbf{0}$, but
$\textsf{E}\left(\hat{\boldsymbol\tau}\right)=\mathbf{Q}_{0}\textsf{E}\left(\hat{\boldsymbol\gamma}\right)+\mathbf{Q}_{1}=\mathbf{Q}_{0}\mathbf{X}_{\textsf{A}}\boldsymbol{\beta} + \mathbf{Q}_{1}$ and $\textsf{E}\left(\boldsymbol\tau\right)=\mathbf{L}\boldsymbol{\beta}$. Then,
$$
\mathbf{0} = \textsf{E}\left(\hat{\boldsymbol\tau}-\boldsymbol\tau\right)
= \mathbf{Q}_{0}\textsf{E}\left(\hat{\boldsymbol\gamma}\right)+\mathbf{Q}_{1} - \mathbf{L}\boldsymbol{\beta}
= \left(\mathbf{Q}_1\mathbf{X}_{\textsf{A}}-\mathbf{L}\right)\boldsymbol{\beta}+\mathbf{Q}_{0}\,,
$$
which requires that both $\mathbf{Q}_{0}=\boldsymbol{0}$ and $\mathbf{Q}_{1}\mathbf{X}_{\textsf{A}} = \mathbf{L}$ hold for the previous chain of equalities to be satisfy.

Thus, the best predictor $\hat{\tau}$ can be found by minimizing $\textsf{Var}\left(\hat{\boldsymbol\tau}-\boldsymbol\tau\right)$ subject to $\mathbf{Q}_{1}\mathbf{X}_{\textsf{A}} = \mathbf{L}$. Thus,
\begin{eqnarray}
\textsf{Var}\left(\hat{\boldsymbol\tau}-\boldsymbol\tau\right) & = & \textsf{Var}\left(\mathbf{Q}_{1}\hat{\boldsymbol\gamma}-\mathbf{L}\boldsymbol{\beta}-\mathbf{M}\boldsymbol{u}\right)\nonumber \\
 & = & \textsf{Var}\left(\mathbf{Q}_{1}\hat{\boldsymbol\gamma}\right)  + \textsf{Var}\left(\mathbf{M}\boldsymbol{u}\right)-2\textsf{Cov}\left(\mathbf{Q}_{1}\hat{\boldsymbol\gamma},\mathbf{M}\boldsymbol{u}\right)\nonumber \\
 & = & \mathbf{Q}_{1}\mathbf{V}\mathbf{Q}_{1}^{\textsf{T}}+\mathbf{M}\mathbf{V}_{u}\mathbf{M}^{\textsf{T}}-2\mathbf{Q}_{1}\mathbf{C}\mathbf{M}^{\textsf{T}}\,,\label{eq:varianza del sesgo}
\end{eqnarray}
with $\mathbf{C} = \textsf{Cov}\left(\hat{\boldsymbol\gamma},\,\boldsymbol{u}\right)$.
Since $\mathbf{M}\mathbf{V}_{u}\mathbf{M}^{\textsf{T}}$ does not
depend on $\mathbf{Q}_{1}$, the minimization problem given above can be restated  as
minimize $\mathbf{Q}_{1}\mathbf{V}\mathbf{Q}_{1}^{\textsf{T}}-2\mathbf{Q}_{1}\mathbf{C}\mathbf{M}^{\textsf{T}}$ subject to $\mathbf{Q}_{1}\mathbf{X}_{\textsf{A}}=\mathbf{L}$,
whose corresponding Lagrangian function is
$$
\ell\left(\mathbf{Q}_{1},\,\boldsymbol{\Lambda}\right) = \mathbf{Q}_{1}\mathbf{V}\mathbf{Q}_{1}^{\textsf{T}}-2\mathbf{Q}_{1}\mathbf{C}\mathbf{M}^{\textsf{T}}+2\left(\mathbf{Q}_{1}\mathbf{X}_{\textsf{A}}-\mathbf{L}\right)\boldsymbol{\Lambda}\,.
$$
Taking partial derivatives with respect to both $\mathbf{Q}_{1}$ and $\boldsymbol{\Lambda}$, we get that
\begin{equation*}
\frac{\partial \ell\left(\mathbf{Q}_{1},\,\boldsymbol{\Lambda}\right)}{\partial\mathbf{Q}_{1}}=2\mathbf{V}\mathbf{Q}_{1}^{\textsf{T}}-2\mathbf{C}\mathbf{M}^{\textsf{T}}+2\mathbf{X}_{\textsf{A}}\boldsymbol{\Lambda} = \mathbf{0}
\qquad\Rightarrow \qquad
\mathbf{V}\mathbf{Q}_{1}^{\textsf{T}}+\mathbf{X}_{\textsf{A}}\boldsymbol{\Lambda}=\mathbf{C}\mathbf{M}^{\textsf{T}}\,,
\end{equation*}
and
\begin{equation*}
\frac{\partial \ell\left(\mathbf{Q}_{1},\,\boldsymbol{\Lambda}\right)}{\partial\text{\ensuremath{\Lambda}}}=2\left(\mathbf{Q}_{1}\mathbf{X}_{\textsf{A}}-\mathbf{L}\right) = \mathbf{0}
\qquad\Rightarrow \qquad
\mathbf{X}_{\textsf{A}}^{\textsf{T}}\mathbf{Q}_{1}^{\textsf{T}}=\mathbf{L}^{\textsf{T}}\,,
\end{equation*}
that in turn can be rewritten in matrix form as
\begin{eqnarray*}
\left(\begin{array}{cc}
\mathbf{V} & \mathbf{X}_{\textsf{A}}\\
\mathbf{X}_{\textsf{A}}^{\textsf{T}} & \boldsymbol{0}
\end{array}\right)\left(\begin{array}{c}
\mathbf{Q}_{1}^{\textsf{T}}\\
\boldsymbol{\Lambda}
\end{array}\right) 
& = & \left(\begin{array}{c}
\mathbf{C}\mathbf{M}^{\textsf{T}}\\
\mathbf{L}^{\textsf{T}}
\end{array}\right)\\
\Rightarrow\left(\begin{array}{c}
\mathbf{Q}_{1}^{\textsf{T}}\\
\boldsymbol{\text{\ensuremath{\Lambda}}}
\end{array}\right) & = & \left(\begin{array}{cc}
\mathbf{V} & \mathbf{X}_{\textsf{A}}\\
\mathbf{X}_{\textsf{A}}^{\textsf{T}} & \boldsymbol{0}
\end{array}\right)^{-1}\left(\begin{array}{c}
\mathbf{C}\mathbf{M}^{\textsf{T}}\\
\mathbf{L}^{\textsf{T}}
\end{array}\right)\,.
\end{eqnarray*}
In this way, 
using standard results from matrix algebra and letting $\mathbf{G}=\left(\mathbf{X}_{\textsf{A}}^{\textsf{T}}\mathbf{V}\mathbf{X}_{\textsf{A}}\right)^{-1}$, we obtain that
\begin{eqnarray*}
\left(\begin{array}{c}
\mathbf{Q}_{1}^{\textsf{T}}\\
\boldsymbol{\text{\ensuremath{\Lambda}}}
\end{array}\right) & = & \left(\begin{array}{cc}
\mathbf{V}^{-1}-\mathbf{V}^{-1}\mathbf{X}_{\textsf{A}}\mathbf{G}\mathbf{X}_{\textsf{A}}^{\textsf{T}}\mathbf{V}^{-1} & \mathbf{V}^{-1}\mathbf{X}_{\textsf{A}}\mathbf{G}\\
\mathbf{G}\mathbf{X}_{\textsf{A}}^{\textsf{T}}\mathbf{V} & -\mathbf{G}
\end{array}\right)\left(\begin{array}{c}
\mathbf{C}\mathbf{M}^{\textsf{T}}\\
\text{\ensuremath{\mathbf{L}^{\textsf{T}}}}
\end{array}\right)\,,
\end{eqnarray*}
which means that
\begin{eqnarray}\label{eq:Q1}
\mathbf{Q}_{1} & = & \left[\mathbf{V}^{-1}\mathbf{X}_{\textsf{A}}\mathbf{G}\mathbf{L}^{\textsf{T}} + \mathbf{V}^{-1}\left(\mathbf{I}-\mathbf{X}_{\textsf{A}}\mathbf{G}\mathbf{X}_{\textsf{A}}^{\textsf{T}}\mathbf{V}^{-1}\right)\mathbf{C}\mathbf{M}^{\textsf{T}}\right]^{\textsf{T}} \nonumber\\ 
& = & \mathbf{L}\mathbf{G}\mathbf{X}_{\textsf{A}}^{\textsf{T}}\mathbf{V}^{-1}+\mathbf{M}\mathbf{C}^{\textsf{T}}\mathbf{V}^{-1}\left(\mathbf{I}-\mathbf{X}_{\textsf{A}}\mathbf{G}\mathbf{X}_{\textsf{A}}^{\textsf{T}}\mathbf{V}^{-1}\right)\,,
\end{eqnarray}
where $\mathbf{I}$ is the identity matrix.

Recall from our earlier discussion that the linear predictor takes the form $\hat{\boldsymbol\tau}=\mathbf{Q}_{1}\hat{\boldsymbol\gamma}$ because it is assumed to be unbiased from the beginning. Thus, substituting $\mathbf{Q}_{1}$ by the expression found in Eq. \eqref{eq:Q1}, we get that the Best Linear Unbiased Prediction (BLUP) of $\boldsymbol\tau$ is given by
\begin{eqnarray*}
\hat{\boldsymbol\tau} & = & \left[\mathbf{L}\mathbf{G}\mathbf{X}_{\textsf{A}}^{\textsf{T}}\mathbf{V}^{-1}+\mathbf{M}\mathbf{C}^{\textsf{T}}\mathbf{V}^{-1}\left(\mathbf{I}-\mathbf{X}_{\textsf{A}}\mathbf{G}\mathbf{X}_{\textsf{A}}^{\textsf{T}}\mathbf{V}^{-1}\right)\right]\hat{\boldsymbol\gamma}\nonumber \\
& = & \mathbf{L}\mathbf{G}\mathbf{X}_{\textsf{A}}^{\textsf{T}}\mathbf{V}^{-1}\hat{\boldsymbol\gamma}+\mathbf{M}\mathbf{C}^{\textsf{T}}\mathbf{V}^{-1}\hat{\boldsymbol\gamma}-\mathbf{M}\mathbf{C}^{\textsf{T}}\mathbf{V}^{-1}\mathbf{X}_{\textsf{A}}\mathbf{G}\mathbf{X}_{\textsf{A}}^{\textsf{T}}\mathbf{V}^{-1}\hat{\boldsymbol\gamma}\nonumber \\
& = & \mathbf{L}\mathbf{G}\mathbf{X}_{\textsf{A}}^{\textsf{T}}\mathbf{V}^{-1}\hat{\boldsymbol\gamma}+\mathbf{M}\mathbf{C}^{\textsf{T}}\mathbf{V}^{-1}\left(\hat{\boldsymbol\gamma}-\mathbf{X}_{\textsf{A}}\hat{\boldsymbol{\beta}}\right)\\
& = & \mathbf{L}\hat{\boldsymbol{\beta}}+\mathbf{M}\hat{\boldsymbol{u}}\,,
\end{eqnarray*}
where $\hat{\boldsymbol{\beta}} = \mathbf{G}\mathbf{X}_{\textsf{A}}^{\textsf{T}}\mathbf{V}^{-1}\hat{\boldsymbol\gamma}$ and $\hat{\boldsymbol{u}}=\mathbf{C}^{\textsf{T}}\mathbf{V}^{-1}\left(\hat{\boldsymbol\gamma}-\mathbf{X}_{\textsf{A}}\hat{\boldsymbol{\beta}}\right)$.

Now, our goal is to see through the previous expression in order to adapt the Fay-Herriot model,
\begin{equation}\label{modeloFH}
\hat{\gamma}_{d} = \boldsymbol{x}_{\textsf{A}\,d}^{\textsf{T}}\,\boldsymbol{\beta}+ u_{d}+e_{d}\,,\qquad d = 1,\ldots,D\,,
\end{equation}
where $\boldsymbol{x}_{\textsf{A}\,d}^{\textsf{T}}\,\boldsymbol{\beta} = \sum_{k=1}^p x_{\textsf{A}\,dk}\,\beta_k$ is a linear predictor of fixed effects and $u_{d}$ is a domain-specific random effect. As usual, all the random effects are assumed to be independent and identically distributed with zero mean and variance $\sigma_{u}^{2}$. 
Furthermore, $e_{d}$ is the sampling error associated with the sampling design $p(\cdot)$, which is also assumed to be independent of $u_{d}$, and also, such that $\textsf{E}_{p}[e_{d}\mid\hat{\gamma}_{d}]=0$ and $\textsf{Var}_{p}[e_{d}\mid\hat{\gamma}_{d}]=\sigma_{d}^{2}$. As a final remark, each $\sigma_{d}^{2}$ can be calculated in a straightforward fashion using Eq. \eqref{Estimadores_directos_y_varianzas}.

Using the results provided in this section and rewriting model \eqref{modeloFH} in matrix form as in Eq. \eqref{eq_LMM}, it follows that
$\hat{\boldsymbol\gamma} = \mathbf{X}_{\textsf{A}}\boldsymbol{\beta}+\mathbf{Z}\boldsymbol{u}+\boldsymbol{e}$
is equivalent to
$$
\left(\begin{array}{c}
\hat{\gamma}_{1}\\
\vdots\\
\hat{\gamma}_{D}
\end{array}\right) = \left(\begin{array}{ccc}
x_{\textsf{A}\,11} & \cdots & x_{\textsf{A}\,1p}\\
\vdots & \ddots & \vdots\\
x_{\textsf{A}\,D1} & \cdots & x_{\textsf{A}\,Dp}
\end{array}\right)\left(\begin{array}{c}
\beta_{1}\\
\vdots\\
\beta_{p}
\end{array}\right)+\left(\begin{array}{c}
u_{1}\\
\vdots\\
u_{D}
\end{array}\right)+\left(\begin{array}{c}
e_{1}\\
\vdots\\
e_{d}
\end{array}\right)\,,
$$
which means that 
$\mathbf{Z}=\mathbf{I}_{D}$ and
$\mathbf{V}=\textsf{diag}\left(\sigma_{u}^{2}+\sigma_{1}^{2},\ldots,\sigma_{u}^{2}+\sigma_{D}^{2}\right)$.
As a consequence, we have that the Best Linear Unbiased Estimator (BLUE) of  $\boldsymbol{\beta}=\left(\beta_{1},\ldots,\beta_{p}\right)$ and $\boldsymbol{u}=\left(u_{1},\ldots,u_{D}\right)$ are respectively
\begin{equation}\label{eq:est_beta_sigma_u}
\tilde{\boldsymbol{\beta}} = \left(\mathbf{X}_{\textsf{A}}^{\textsf{T}}\mathbf{V}^{-1}\mathbf{X}_{\textsf{A}}\right)^{-1}\mathbf{X}_{\textsf{A}}^{\textsf{T}}\mathbf{V}\hat{\boldsymbol{\gamma}}
\qquad\text{and}\qquad
\tilde{\boldsymbol{u}} = \mathbf{C}^{\textsf{T}}\mathbf{V}^{-1}\left(\hat{\boldsymbol{\gamma}}-\mathbf{X}_{\textsf{A}}\tilde{\boldsymbol{\beta}}\right)\,,
\end{equation}
with $\mathbf{V}^{-1}=\textsf{diag}\left(1/\left(\sigma_{u}^{2}+\sigma_{1}^{2}\right),\ldots,1/\left(\sigma_{u}^{2}+\sigma_{D}^{2}\right)\right)$ and $\hat{\boldsymbol{\gamma}}=(\hat\gamma_1,\ldots,\hat\gamma_D)$.

Therefore, the BLUP proposed in this article is
\begin{equation}\label{Mi_BLUP}
\hat{\gamma}_{d}^{\textsf{B}}=\frac{\sigma_{u}^{2}}{\sigma_{u}^{2}+\sigma_{d}^{2}}\,\hat{\gamma}_{d}+\frac{\sigma_{d}^{2}}{\sigma_{u}^{2}+\sigma_{d}^{2}}\,\boldsymbol{x}_{\textsf{A}\,d}^{\textsf{T}}\,\tilde{\boldsymbol{\beta}}\,.
\end{equation}
Also, the Empirical Best Linear Unbiased Predictor (EBLUP) of the mean of the small area $\gamma_{d}$ under model \eqref{modeloFH} can be obtained just by replacing $\sigma_{u}^{2}$ by its corresponding estimate $\hat{\sigma}_{u}^{2}$ in Eq. \eqref{Mi_BLUP}, i.e.,
\begin{equation}\label{eq:estimador_propuesto}
\hat{\gamma}_{d}^{\textsf{P}} = \frac{\hat{\sigma}_{u}^{2}}{\hat{\sigma}_{u}^{2}+\sigma_{d}^{2}}\,\hat{\gamma}_{d}+\frac{\sigma_{d}^{2}}{\hat{\sigma}_{u}^{2}+\sigma_{d}^{2}}\,\boldsymbol{x}_{\textsf{A}\,d}^{\textsf{T}}\,\tilde{\boldsymbol{\beta}}
 = \left(1-B_{d}\right)\hat{\gamma}_{d}+B_{d}\,\boldsymbol{x}_{\textsf{A}\,d}^{\textsf{T}}\,\tilde{\boldsymbol{\beta}}\,,
\end{equation}
with $B_{d}=\sigma_{d}^{2}/(\hat{\sigma}_{u}^{2}+\sigma_{d}^{2})$.

The proposed estimator $\hat{\gamma}_{d}^{\textsf{P}}$ is unbiased for $\gamma_{d}$, for $d=1,\ldots,D$. 
Indeed, consider the difference
\begin{eqnarray*}
\hat{\gamma}_{d}^{\textsf{P}}-\gamma_{d} & = & \left(1-B_{d}\right)\hat{\gamma}_{d}+B_{d}\,\boldsymbol{x}_{\textsf{A}\,d}^{\textsf{T}}\,\tilde{\boldsymbol{\beta}} - \left[B_{d}\,\gamma_{d}+\left(1-B_{d}\right)\gamma_{d}\right].
\end{eqnarray*}
Now, by letting $\alpha_{d}=1-B_{d}$, $\gamma_{d}=\boldsymbol{x}_{\textsf{A}\,d}^{\textsf{T}}\,\boldsymbol{\beta} + u_{d}$, and $\hat{\gamma}_{d}=\gamma_{d}+e_{d}$, we have that
\begin{eqnarray*}
\hat{\gamma}_{d}^{\textsf{P}}-\gamma_{d} & = & \alpha_{d}\left(\gamma_{d}+e_{d}\right)+\left(1-\alpha_{d}\right)\boldsymbol{x}_{\textsf{A}\,d}^{\textsf{T}}\,\tilde{\boldsymbol{\beta}}-\alpha_{d}\gamma_{d}-\left(1-\alpha_{d}\right)\gamma_{d}\\
 & = & \alpha_{d}e_{d}+\left(1-\alpha_{d}\right)\boldsymbol{x}_{\textsf{A}\,d}^{\textsf{T}}\,\tilde{\boldsymbol{\beta}}-\left(1-\alpha_{d}\right)\left[\boldsymbol{x}_{\textsf{A}\,d}^{\textsf{T}}\,\boldsymbol{\beta}+u_{d}\right]\\
 & = & \alpha_{d}e_{d}+\left(1-\alpha_{d}\right)\left[\boldsymbol{x}_{\textsf{A}\,d}^{\textsf{T}}\,(\tilde{\boldsymbol{\beta}}-\boldsymbol{\beta}  )\right]-\left(1-\alpha_{d}\right)u_{d}\,.
\end{eqnarray*}
Then, taking expected values,
\begin{eqnarray*}
\textsf{E}\left(\hat{\gamma}_{d}^{\textsf{P}}-\gamma_{d}\right) = \textsf{E}\left(\alpha_{d}e_{d}+\left(1-\alpha_{d}\right)\left[\boldsymbol{x}_{\textsf{A}\,d}^{\textsf{T}}\,(\tilde{\boldsymbol{\beta}}-\boldsymbol{\beta})\right]-\left(1-\alpha_{d}\right)u_{d}\right) = 0\,,
\end{eqnarray*}
because $\textsf{E}\left(e_{d}\right) = \textsf{E}\left(u_{d}\right)=0$, and also, $\textsf{E}(\tilde{\boldsymbol{\beta}}-\boldsymbol{\beta})=\boldsymbol{0}$
since $\tilde{\boldsymbol{\beta}}$ is an unbiased estimator for $\boldsymbol{\beta}$.

\subsection{Mean square error of the proposed estimator}\label{sec_MSE}

Here, we follow very closely \cite{kackar1984approximations}, \citet{prasad1990estimation} as well as \citet{ghosh1994small}, in order to obtain an expression for the  Mean Square Error (MSE) of the proposed estimator $\hat{\gamma}_{d}^{\textsf{P}}$.
Specifically, we consider the variance component estimation method, which require us to find three quantities explicitly, namely, $g_{1d}(\sigma_{u}^{2})$, $g_{2d}(\sigma_{u}^{2})$, and $g_{3d}(\sigma_{u}^{2})$, for $d=1,\ldots,D$.
We refer the reader to the previous references for details about such a method. However, we outline below some fundamental details.

First, in order to calculate $g_{1d}\left(\sigma_{u}^{2}\right)$, we need to tak into account that
$\mathbf{V}_{u}=\sigma_{u}^{2}\mathbf{I}_{D}$
and
$\mathbf{V}_{e}=\sigma_{d}^{2}\mathbf{W}_{N}^{-1}\mathbf{V}_{s}^{-1}=\textsf{diag}\left(\mathbf{V}_{s_1}^{-1},\ldots,\mathbf{V}_{s_D}^{-1}\right)$,
where $\mathbf{W}_{N}$ is a $N\times N$ diagonal matrix of weights induced by the sampling design $p(\cdot)$, with $N$ the population size, $s=\cup_{d=1}^D s_d$ is the full sample under consideration, with $s_d$ the sample in domain $d$, and
$$
\mathbf{V}_{s_d}^{-1}=\frac{1}{\sigma_{d}^{2}}\left(\mathbf{W}_{s_d}-\frac{B_{d}}{w_{d}}\,\boldsymbol{w}_{n_d}\boldsymbol{w}_{n_d}^{\textsf{T}}\right)\,,
$$
where $w_d$ is the weight of domain $d$, for $d=1,\ldots,D$. Thus,
\begin{eqnarray*}
\mathbf{V}_{u}\mathbf{Z}_{s}^{\textsf{T}}\mathbf{V}_{e\,s}^{-1}\mathbf{Z}_{s}\mathbf{V}_{u} 
& = & \frac{\sigma_{u}^{4}}{\sigma_{d}^{2}}\,\textsf{diag}\left(\boldsymbol{1}_{n_{1}}^{\textsf{T}},\ldots,\boldsymbol{1}_{n_{D}}^{\textsf{T}}\right)\,\cdot\\
&   &\qquad\,\,\,\,\,\,\, \textsf{diag}\left(\mathbf{W}_{n_{1}}-\frac{B_{1}}{n_{1}}\boldsymbol{w}_{n_{1}}\boldsymbol{1}_{w_{1}}^{\textsf{T}},\ldots,\mathbf{W}_{n_{D}}-\frac{B_{D}}{n_{D}}\boldsymbol{w}_{n_{D}}\boldsymbol{1}_{D}^{\textsf{T}}\right)\,\cdot\\
 &  & \qquad\qquad\,\,\,\,\,\,\,\textsf{diag}\left(\boldsymbol{1}_{n_{1}},\ldots,\boldsymbol{1}_{n_{D}}\right)\\
 & = & \sigma_{u}^{2}\,\textsf{diag}\left(\frac{B_{1}}{w_{1}}\boldsymbol{w}_{n_{1}}^{\textsf{T}},\ldots,\frac{B_{D}}{w_{D}}\boldsymbol{w}_{n_{D}}^{\textsf{T}}\right)\,\textsf{diag}\left(\boldsymbol{1}_{n_{1}},\ldots,\,\boldsymbol{1}_{n_{D}}\right)\\
 & = & \sigma_{u}^{2}\,\textsf{diag}\left(B_{1},\ldots,B_{D}\right)\,,
\end{eqnarray*}
and therefore,
\begin{eqnarray*}
\mathbf{T}_{s} = \mathbf{V}_{u}(\mathbf{I}_D-\mathbf{Z}_{s}^{\textsf{T}}\mathbf{V}_{e\,s}^{-1}\mathbf{Z}_{s}\mathbf{V}_{u})  = \sigma_{u}^{2}\,\textsf{diag}\left(1-B_{1},\ldots,1-B_{D}\right)\,,
\end{eqnarray*}
with $\mathbf{Z}_s = \textsf{diag}\left(\boldsymbol{1}_{n_{1}},\ldots,\boldsymbol{1}_{n_{D}}\right)$ and 
$$
\mathbf{V}_{e\,s} = \textsf{diag}\left(\mathbf{W}_{n_{1}}-\frac{B_{1}}{n_{1}}\boldsymbol{w}_{n_{1}}\boldsymbol{1}_{w_{1}}^{\textsf{T}},\ldots,\mathbf{W}_{n_{D}}-\frac{B_{D}}{n_{D}}\boldsymbol{w}_{n_{D}}\boldsymbol{1}_{D}^{\textsf{T}}\right)\,.
$$
The previous result is quite useful because we want to estimate the average of the plausible values for all the individuals within domain $d$, i.e., $\eta = \boldsymbol{a}^{\textsf{T}}\boldsymbol{\gamma}_{\textsf{U}}$, with
\begin{align*}
\boldsymbol{a}^{\textsf{T}} & =\frac{1}{N_{d}}\left(\boldsymbol{0}_{N_{1}}^{\textsf{T}},\ldots,\boldsymbol{0}_{N_{d-1}}^{\textsf{T}},\boldsymbol{1}_{N_{d}}^{\textsf{T}},\boldsymbol{0}_{N_{d+1}}^{\textsf{T}},\ldots,\boldsymbol{0}_{N_{D}}^{\textsf{T}}\right)\,,
\end{align*}
with $\boldsymbol{\gamma}_{\textsf{U}}$ is the population parameter, leading us directly to consider all those individuals included the sample as well as those that did not, i.e, $s$ and $r$, respectively. As a consequence, 
\begin{eqnarray} \label{Estimacion_g1_metodologia}
g_{1d}\left(\sigma_{u}^{2}\right) & = & \boldsymbol{a}_{r}^{\textsf{T}}\mathbf{Z}_{r}\mathbf{T}_{s}\mathbf{Z}_{r}^{\textsf{T}}\boldsymbol{a}_{r}\nonumber \\
& = & \left(\boldsymbol{0}^{\textsf{T}},\ldots,\boldsymbol{0}^{\textsf{T}},\boldsymbol{1}_{N_{d}-n_{d}}^{\textsf{T}},\boldsymbol{0}^{\textsf{T}},\ldots,\boldsymbol{0}^{\textsf{T}}\right)\cdot\textsf{diag}\left(\boldsymbol{1}_{N_{1}-n_{1}},\ldots,\boldsymbol{1}_{N_{d}-n_{d}}\right)\cdot\nonumber\\ 
&  & \qquad\,\,\,\,\frac{\sigma_{u}^{2}}{N_{d}^{2}}\,\textsf{diag}\left(1-B_{1},\ldots,1-B_{D}\right)\cdot\nonumber \\
&  & \qquad\qquad\,\,\,\,\,\,\textsf{diag}\left(\boldsymbol{1}_{N_{1}-n_{1}}^{\textsf{T}},\ldots,\boldsymbol{1}_{N_{D}-n_{D}}^{\textsf{T}}\right)\left(\boldsymbol{0}^{\textsf{T}},\ldots,\boldsymbol{0}^{\textsf{T}},\boldsymbol{1}_{N_{d}-n_{d}}^{\textsf{T}},\boldsymbol{0}^{\textsf{T}},\ldots,\boldsymbol{0}^{\textsf{T}}\right)\nonumber\\
& = & \frac{\sigma_{u}^{2}}{N_{d}^{2}}\left(1-B_{d}\right)\left(N_{d}-n_{d}\right)^{2}\nonumber \\
& \simeq & \sigma_{u}^{2}\left(1-B_{d}\right)\,, 
\end{eqnarray}
for $n_{d} << N_{d}$.

Now, in order to compute $g_{2d}\left(\sigma_{u}^{2}\right)$, we need to get
\begin{eqnarray*}
\mathbf{Z}_{r}\mathbf{T}_{s}\mathbf{Z}_{s}^{\textsf{T}} & = & \,\textsf{diag}\left(\boldsymbol{1}_{N_{1}-n_{1}},\ldots,\boldsymbol{1}_{N_{D}-n_{D}}\right)\cdot\\
& &\qquad\,\,\,\,\sigma_{u}^{2}\,\textsf{diag}\left(1-B_{1},\ldots,1-B_{D}\right)\cdot\\
& &\qquad\qquad\,\,\,\,\,\,\textsf{diag}\left(\boldsymbol{1}_{n_{1}}^{\textsf{T}},\ldots,\boldsymbol{1}_{n_{D}}^{\textsf{T}}\right) \\
& = & \sigma_{u}^{2}\,\textsf{diag}\left(\left(1-B_{1}\right)\,1_{N_{1}-n_{1}}\boldsymbol{1}_{n_{1}}^{\textsf{T}},\ldots,\left(1-B_{D}\right)\,\boldsymbol{1}_{N_{D}-n_{D}}\boldsymbol{1}_{n_{D}}^{\textsf{T}}\right)\,,
\end{eqnarray*}
which means that,
\begin{eqnarray}\label{eq:-4}
\boldsymbol{a}_{r}^{\textsf{T}}\mathbf{Z}_{r}\mathbf{T}_{s}\mathbf{Z}_{s}^{\textsf{T}}\mathbf{V}_{e\,s}^{-1}\mathbf{X}_{\textsf{A}\,s} & = & \frac{1}{N_{d}}\frac{\sigma_{u}^{2}}{\sigma_{d}^{2}}\left(\boldsymbol{0}^{\textsf{T}},\ldots,\boldsymbol{0}^{\textsf{T}},\boldsymbol{1}_{N_{d}-n_{d}}^{\textsf{T}},\boldsymbol{0}^{\textsf{T}},\ldots,\boldsymbol{0}^{\textsf{T}}\right)\cdot\nonumber \\
&  & \qquad\,\,\,\, \textsf{diag}\left(\left(1-B_{1}\right)\,\boldsymbol{1}_{N_{1}-n_{1}}\boldsymbol{1}_{n_{1}}^{\textsf{T}},\ldots,\left(1-B_{D}\right)\,\boldsymbol{1}_{N_{D}-n_{D}}\boldsymbol{1}_{n_{D}}^{\textsf{T}}\right)\cdot\nonumber\\
&  & \qquad\qquad\,\,\,\,\,\,\mathbf{W}_{s}\mathbf{X}_{\textsf{A}\,s}\nonumber \\
& = & \frac{1}{N_{d}}\frac{\sigma_{u}^{2}}{\sigma_{d}^{2}}\left(1-B_{d}\right)\left(N_{d}-n_{d}\right)\left(\boldsymbol{0}^{\textsf{T}},\ldots,\boldsymbol{0}^{\textsf{T}},\boldsymbol{w}_{n_{d}}^{\textsf{T}},\boldsymbol{0}^{\textsf{T}},\ldots,\boldsymbol{0}^{\textsf{T}}\right)\mathbf{X}_{\textsf{A}\,s}\nonumber \\
& = & \left(1-f_{d}\right)B_{d}\,\hat{\bar{\boldsymbol{x}}}_{d}\,,
\end{eqnarray}
with $f_d=n_d/N_d$ is the sampling fraction and $\hat{\bar{\boldsymbol{x}}}_{d}=\frac{1}{\sum w_{d}}\sum_{k\in s_{d}} x_{dk}\boldsymbol{w}_{dk}$.
Moreover,
\begin{eqnarray}\label{eq:-5}
\boldsymbol{a}_{r}^{\textsf{T}}\mathbf{X}_{\textsf{A}\,r} = \frac{1}{N_{d}}\left(\boldsymbol{0}^{\textsf{T}},\ldots,\boldsymbol{0}^{\textsf{T}},\boldsymbol{1}_{N_{d}-n_{d}}^{\textsf{T}},\boldsymbol{0}^{\textsf{T}},\ldots,\boldsymbol{0}^{\textsf{T}}\right)\mathbf{X}_{\textsf{A}\,r}
= \left(1-f_{d}\right)\bar{\boldsymbol{x}}_{d}\,,
\end{eqnarray}  
with $\bar{\boldsymbol{x}}_{d} = \frac{1}{N_{d}}\sum_{k\in d} x_{dk}$.
Using together Eqs. \eqref{eq:-4} and \eqref{eq:-5}, along with $\ensuremath{n_{d}<<N_{d}}$,
we finally get that
\begin{eqnarray}
g_{2d}\left(\sigma_{u}^{2}\right) & = & \left(\bar{\boldsymbol{x}}_{d}-B_{d}\,\hat{\bar{\boldsymbol{x}}}_{d}\right)\left(\mathbf{X}_{\textsf{A}\,s}^{\textsf{T}}\mathbf{V}_{s}^{-1}\mathbf{X}_{\textsf{A}\,s}\right)^{-1}\left(\bar{\boldsymbol{x}}_{d}-B_{d}\,\hat{\bar{\boldsymbol{x}}}_{d}\right)^{\textsf{T}}\,.\label{g2_metodologia}
\end{eqnarray}

Lastly, in order to $g_{3d}\left(\sigma_{u}^{2}\right)$, we need to get
\begin{eqnarray*}
\boldsymbol{b}^{\textsf{T}} & = & \frac{1}{N_{d}}\frac{\sigma_{u}^{2}}{\sigma_{d}^{2}}\left(\boldsymbol{0}^{\textsf{T}},\ldots,\boldsymbol{0}^{\textsf{T}},\boldsymbol{1}_{N_{d}-n_{d}}^{\textsf{T}},\boldsymbol{0}^{\textsf{T}},\ldots,\boldsymbol{0}^{\textsf{T}}\right)\cdot\nonumber \\
&  & \qquad\,\,\,\,\,\textsf{diag}\left(\boldsymbol{1}_{N_{1}-n_{1}},\ldots,\boldsymbol{1}_{N_{D}-n_{D}}\right)\textsf{diag}\left(\boldsymbol{1}_{n_{1}}^{\textsf{T}},\ldots,\boldsymbol{1}_{n_{D}}^{\textsf{T}}\right)\cdot\nonumber\\
&  & \qquad\qquad\,\,\,\,\,\,\textsf{diag}\left(\mathbf{W}_{n_{1}}-\frac{B_{1}}{w_{1}}\,\boldsymbol{w}_{n_{1}}\boldsymbol{w}_{n_{1}}^{\textsf{T}},\ldots,\mathbf{W}_{n_{D}}-\frac{B_{D}}{w_{D}}\,\boldsymbol{w}_{n_{D}}\boldsymbol{w}_{n_{D}}^{\textsf{T}}\right)\nonumber \\
& = & \frac{1}{N_{d}}\left(\boldsymbol{0}^{\textsf{T}},\ldots,\boldsymbol{0}^{\textsf{T}},\boldsymbol{1}_{N_{d}-n_{d}}^{\textsf{T}},\boldsymbol{0}^{\textsf{T}},\ldots,\boldsymbol{0}^{\textsf{T}}\right)\cdot\nonumber\\ &  & \qquad\,\,\,\,\textsf{diag}\left(\frac{B_{1}}{w_{1}}\boldsymbol{1}_{N_{1}-n_{1}}\boldsymbol{w}_{n_{1}}^{\textsf{T}},\ldots,\frac{B_{D}}{w_{D}}\boldsymbol{1}_{N_{D}-n_{D}}\boldsymbol{w}_{n_{D}}^{\textsf{T}}\right)\nonumber \\
& = & \left(\boldsymbol{0}^{\textsf{T}},\ldots,\boldsymbol{0}^{\textsf{T}},\frac{B_{d}}{w_{d}}\frac{N_{d}-n_{d}}{N_{d}}\,\boldsymbol{w}_{n_{d}}^{\textsf{T}},\boldsymbol{0}^{\textsf{T}},\ldots,\boldsymbol{0}^{\textsf{T}}\right)\,.
\end{eqnarray*}
Then,
\begin{eqnarray*}
\nabla\boldsymbol{b}^{\textsf{T}} & = & 
\left(\begin{array}{c}
\boldsymbol{0}^{\textsf{T}},\ldots,\boldsymbol{0}^{\textsf{T}},\left(1-f_{d}\right)\frac{\partial B_{d}}{\partial\sigma_{d}^{2}}\frac{1}{w_{d}}\,\boldsymbol{w}_{n_{d}}^{\textsf{T}},\boldsymbol{0}^{\textsf{T}},\ldots,\boldsymbol{0}^{\textsf{T}}\\
\boldsymbol{0}^{\textsf{T}},\ldots,\boldsymbol{0}^{\textsf{T}},\left(1-f_{d}\right)\frac{\partial B_{d}}{\partial\sigma_{u}^{2}}\frac{1}{w_{d}}\,\boldsymbol{w}_{n_{d}}^{\textsf{T}},\boldsymbol{0}^{\textsf{T}},\ldots,\boldsymbol{0}^{\textsf{T}}
\end{array}\right)
\end{eqnarray*}
and therefore,
\begin{multline*}
g_{3d}\left(\sigma_{u}^{2}\right) = \left(1-f_{d}\right)^{2}\left(\sigma_{u}^{2}+\frac{\sigma_{d}^{2}}{w_{d}}\right)\cdot\\ \textsf{tr}\left\{ 
\left(\begin{array}{cc}
\left(\frac{\partial B_{d}}{\partial\sigma_{d}^{2}}\right)^{2} & \frac{\partial B_{d}}{\partial\sigma_{d}^{2}}\,\frac{\partial B_{d}}{\partial\sigma_{u}^{2}}\\
\frac{\partial B_{d}}{\partial\sigma_{d}^{2}}\,\frac{\partial B_{d}}{\partial\sigma_{u}^{2}} & \left(\frac{\partial B_{d}}{\partial\sigma_{u}^{2}}\right)^{2}
\end{array}\right)\left(\begin{array}{cc}
\textsf{Var}\left(\hat{\sigma}_{d}^{2}\right) & \textsf{Cov}\left(\hat{\sigma}_{d}^{2},\,\hat{\sigma}_{u}^{2}\right)\\
\textsf{Cov}\left(\hat{\sigma}_{d}^{2},\,\hat{\sigma}_{u}^{2}\right) & \textsf{Var}\left(\hat{\sigma}_{u}^{2}\right)
\end{array}\right)\right\} \nonumber
\end{multline*}
which means that,
\begin{equation}\label{g3_metodologia}
g_{3d}\left(\sigma_{u}^{2}\right) =\left(\sigma_{u}^{2}+\frac{\sigma_{d}^{2}}{w_{d}}\right)^{-3}\,\frac{1}{w_{d}^{2}}\,\textsf{Var}\left(\sigma_{u}^{2}\hat{\sigma}_{d}^{2}-\sigma_{d}^{2}\hat{\sigma}_{u}^{2}\right)
\end{equation}
for $n_{d} << N_{d}$.

The estimation of the MSE for the proposed estimator $\hat{\gamma}_{d}^{\textsf{P}}$ can be now obtained taking into account our findings given in Eqs. \eqref{Estimacion_g1_metodologia}, \eqref{g2_metodologia}, and \eqref{g3_metodologia}. Specifically, considering $\sigma_u^2$ instead of $\hat\sigma_u^2$, it follows that 
$$
g_{1d}\left(\hat{\sigma}_{u}^{2}\right) =\frac{\hat{\sigma}_{u}^{2}\sigma_{d}^{2}}{\hat{\sigma}_{u}^{2}+\sigma_{d}^{2}}
= \sigma_{d}^{2}\,\left(1-B_{d}\right)
\qquad\text{and}\qquad
g_{2d}\left(\hat{\sigma}_{u}^{2}\right) = \left(\frac{\sigma_{d}^{2}}{\hat{\sigma}_{u}^{2}+\sigma_{d}^{2}}\right)^{2}a
= B_{d}^{2}\,a\,,
$$
where  $B_{d}=\sigma_{d}^{2}/(\hat{\sigma}_{u}^{2}+\sigma_{d}^{2})$ and $a = \boldsymbol{x}_{d\,\textsf{A}}\mathbf{F}^{-1}\boldsymbol{x}_{d\,\textsf{A}}^{\textsf{T}}$, with $\mathbf{F} = (\hat{\sigma}_{u}^{2}+\sigma_{d}^{2})^{-1}\,\sum_{d=1}^{D}\boldsymbol{x}_{d\,\textsf{A}}\,\boldsymbol{x}_{d\,\textsf{A}}^{\textsf{T}}$.
The component $g_{3d}\left(\hat{\sigma}_{u}^{2}\right)$ depends on the estimation method of the variance components. Either way, $g_{3d}\left(\hat{\sigma}_{u}^{2}\right)$ takes the form
$$
g_{3d}\left(\hat{\sigma}_{u}^{2}\right) = 
\frac{\sigma_{d}^{4}}{\left(\hat{\sigma}_{u}^{2}+\sigma_{d}^{2}\right)^{3}}\,\textsf{var}\left(\hat{\sigma}_{u}^{2}\right)\,,
$$
where 
$$
\textsf{Var}\left(\hat{\sigma}_{u}^{2}\right) = 
\frac{2}{D}\left[\hat{\sigma}_{u}^{4}+\frac{2\hat{\sigma}_{u}^{2}}{D}\sum_{d=1}^{D}\sigma_{d}^{2}+\frac{1}{D}\sum_{d=1}^{D}\sigma_{d}^{4}\right]
\quad\text{or}\quad
\textsf{Var}\left(\hat{\sigma}_{u}^{2}\right) = 
2\left[\sum_{d=1}^{D}\left(\hat{\sigma}_{u}^{2}+\sigma_{d}^{2}\right)^{-2}\right]^{-1}\,,
$$
using the Prasad-Rao moment estimator \citep{prasad1990estimation}, or
either the Maximum Likelihood (ML) estimator or the Restricted Maximum Likelihood (REML) estimator, respectively. If the latter, $g_{3d}\left(\hat{\sigma}_{u}^{2}\right)$ simplifies to
$$
g_{3d}\left(\hat{\sigma}_{u}^{2}\right) =
\left(\frac{1}{\hat{\sigma}_{u}^{2}+\sigma_{d}^{2}}\right)\,\frac{2\,B_{d}^{2}}{\sum_{d=1}^{D}\left(\hat{\sigma}_{u}^{2}+\sigma_{d}^{2}\right)^{2}}\,.
$$
Finally, on the one hand, if either Prasad-Rao or REML are used, then the MSE estimator is given by
$$
\widehat{\textsf{MSE}}(\,\hat{\gamma}_{d}^{\textsf{P}}\,) = g_{1\,d}\left(\hat{\sigma}_{u}^{2}\right)+g_{2\,d}\left(\hat{\sigma}_{u}^{2}\right)+2g_{3\,d}\left(\hat{\sigma}_{u}^{2}\right)\,,
$$
but on the other, if ML is used, then 
$$
\widehat{\textsf{MSE}}(\,\hat{\gamma}_{d}^{\textsf{P}}\,) =
g_{1\,d}\left(\hat{\sigma}_{u}^{2}\right)+g_{2\,d}\left(\hat{\sigma}_{u}^{2}\right)+2g_{3\,d}\left(\hat{\sigma}_{u}^{2}\right)-b\,\nabla g_{1}\,,
$$
with
\begin{eqnarray*}
b & = & -\left[\sum_{d=1}^{D}\left(\hat{\sigma}_{u}^{2}+\sigma_{d}^{2}\right)^{-2}\right]^{-1}\cdot\\
& & \qquad\,\,\,\,\, \textsf{tr}\left\{ \left(\sum_{d=1}^{D}\left(\hat{\sigma}_{u}^{2}+\sigma_{d}^{2}\right)^{-1}\boldsymbol{x}_{d\,\textsf{A}}\,\boldsymbol{x}_{d\,\textsf{A}}^{\textsf{T}}\right)^{-1}\left(\sum_{d=1}^{D}\left(\hat{\sigma}_{u}^{2}+\sigma_{d}^{2}\right)^{-2}\boldsymbol{x}_{d\,\textsf{A}}\,\boldsymbol{x}_{d\,\textsf{A}}^{\textsf{T}}\right)\right\}
\end{eqnarray*}
and $\nabla g_{1} = \sigma_{d}^{4}\left(\hat{\sigma}_{u}^{2}+\sigma_{d}^{2}\right)^{-2}$.

\section{Simulation study}\label{sec_simulation_study}

In this section, we carry out a simulation study in order to assess the statistical properties of the proposed estimators. The aim of this simulation is to analyze the Relative Bias,
$SB_{d}=\frac{\gamma_{d}-\hat{\gamma}_{d}}{\hat{\gamma}_{d}}$,
the Relative Standard Error $EER_{d}=\frac{RMSE(\hat{\gamma}_{d})}{\gamma_{d}}\times100\%$,
as well as the Relative Mean Square Error $RMSE\left(\hat{\gamma}_{d}\right)=\sqrt{MSE\left(\hat{\gamma}_{d}\right)}$
associated with the proposed estimator versus the Horvitz-Thompson estimator (\citealt{narain1951sampling}, \citealt{horvitz1952generalization}), the calibration estimator, and the composite estimator.

Our simulation study follows the following steps:
\begin{enumerate}
\item Simulate a population of $N=100,000$ students with 150 items per student and $D=500$ domains, along with two auxiliary variables associated with the ability $\theta$ for each student.
\item Set 10\%, 20\% and 30\% of missing responses per student completely at random in the population. 
\item Define two (2) auxiliary variables at the domain level correlated with $\theta_{d}$ on three levels
(high $\left(>80\%\right)$, medium $\left(60\%\,,\,80\%\right)$ and low $\left(<60\%\right)$). 
\item Estimate five (5) plausible values for each student according to Equation {\color{red} ??}. 
\item Consider a different number of domains in the sample using different sampling fractions as $f_{d}=30\%,\,50\%,\,\text{and}\,70\%$ on the total number of domains in the population.
\item For the domains selected in step 5., select a random sample from the population using simple random sampling with sampling fraction $f_{n}=5\%,\,10\%,\text{and}\,\,20\%$, considering the Horvitz-Thompson estimate, $\hat{\gamma}_{d}^{\textsf{Dir}}$, the calibration estimate, $\hat{\gamma}_{d}^{\textsf{Cal}}$, and the composite estimator, $\hat{\gamma}_{d}^{\textsf{Comp}}$, the proposed estimator, $\hat{\gamma}_{d}^{\textsf{P}}$ using the REML method, 
for each selected domain, along with its standard deviation.
\item Calculate the relative biases, $SB_{d},$ and the relative standard errors, $EER_{d}$, by domain, for each of the estimators $\hat{\gamma}_{d}^{\textsf{Dir}}$, $\hat{\gamma}_{d}^{\textsf{Cal}}$, $\hat{\gamma}_{d}^{\textsf{Comp}}$, and $\hat{\gamma}_{d}^{\textsf{P}}$, and compute: 
\begin{enumerate}
\item $SBP=\frac{{\sum_{d=1}^{D}SB_{d}}}{D}$ (average relative bias). 
\item $EERP=\frac{{\sum_{d=1}^{D}EERP_{d}}}{D}$ (average relative standard error).
\end{enumerate}
\item Repeat steps 5. to 7. for 100,000 random samples and compute:
\begin{enumerate}
\item $\overline{SBR}=\frac{{\displaystyle \sum_{r=1}^{100000}SBP_{r}}}{100000}$
(mean of the average relative biases). 
\item $\overline{EERP}=\frac{{\displaystyle \sum_{r=1}^{100000}EERP_{r}}}{100000}$
(mean of the average relative standard error).
\end{enumerate}
\end{enumerate}

\subsection{Simulation Study Results}

\begin{table}[H]
\begin{centering}
{\scriptsize{}}%
\begin{tabular}{ccccccc}
\hline 
{\scriptsize{}$f_{d}$ (\%) } & {\scriptsize{}$f_{n}$(\%) } & {\scriptsize{}$\overline{EERP}$$\hat{\gamma}_{d}^{\textsf{Dir}}$(\%)} & {\scriptsize{}$\overline{EERP}$$\hat{\gamma}_{d}^{\textsf{Cal}}$(\%)} & {\scriptsize{}$\overline{EERP}$$\hat{\gamma}_{d}^{\textsf{Comp}}$ (\%)} & {\scriptsize{}$\overline{EERP}$$\hat{\gamma}_{d}^{\textsf{P}}$(\%)} & {\scriptsize{}$\overline{SBR}$$\hat{\gamma}_{d}^{\textsf{P}}$ (\%)}\tabularnewline
\hline 
{\scriptsize{}30\%} & {\scriptsize{}5\%} & {\scriptsize{}1.53} & {\scriptsize{}1.20} & {\scriptsize{}1.03} & {\scriptsize{}0.85} & {\scriptsize{}0.08}\tabularnewline
{\scriptsize{}30\%} & {\scriptsize{}10\%} & {\scriptsize{}1.23} & {\scriptsize{}1.07} & {\scriptsize{}1.00} & {\scriptsize{}0.85} & {\scriptsize{}0.02}\tabularnewline
{\scriptsize{}30\%} & {\scriptsize{}20\%} & {\scriptsize{}1.03} & {\scriptsize{}1.00} & {\scriptsize{}0.98} & {\scriptsize{}0.85} & {\scriptsize{}-0.01}\tabularnewline
{\scriptsize{}50\%} & {\scriptsize{}5\%} & {\scriptsize{}1.87} & {\scriptsize{}1.38} & {\scriptsize{}1.09} & {\scriptsize{}0.88} & {\scriptsize{}0.14}\tabularnewline
{\scriptsize{}50\%} & {\scriptsize{}10\%} & {\scriptsize{}1.45} & {\scriptsize{}1.16} & {\scriptsize{}1.02} & {\scriptsize{}0.86} & {\scriptsize{}0.05}\tabularnewline
{\scriptsize{}50\%} & {\scriptsize{}20\%} & {\scriptsize{}1.17} & {\scriptsize{}1.04} & {\scriptsize{}0.99} & {\scriptsize{}0.87} & {\scriptsize{}0.01}\tabularnewline
{\scriptsize{}70\%} & {\scriptsize{}5\%} & {\scriptsize{}2.16} & {\scriptsize{}1.56} & {\scriptsize{}1.15} & {\scriptsize{}0.96} & {\scriptsize{}0.22}\tabularnewline
{\scriptsize{}70\%} & {\scriptsize{}10\%} & {\scriptsize{}1.63} & {\scriptsize{}1.24} & {\scriptsize{}1.05} & {\scriptsize{}0.87} & {\scriptsize{}0.09}\tabularnewline
{\scriptsize{}70\%} & {\scriptsize{}20\%} & {\scriptsize{}1.29} & {\scriptsize{}1.09} & {\scriptsize{}1.01} & {\scriptsize{}0.89} & {\scriptsize{}0.03}\tabularnewline
\hline 
\end{tabular}{\scriptsize\par}
\par\end{centering}
\caption{Missing 10\% and high correlation \label{tabla_1_media}}
\end{table}

Table \ref{tabla_1_media} considers the simulation scenario when the percentage of missing values is 10\% and there is a high correlation between the auxiliary variables and the mean ability. Our findings make evident that the proposed estimator is unbiased as it was theoretically shown. 
In all the scenarios, the proposed estimator has a lower $\overline{EERP}$ as well as a lower mean squared error compared to the alternative estimators. 
One of the scenarios where the $\overline{EERP}$ is higher for all the considered estimators is when the percentage of domains is 70\% (350 domains) and sample fraction of 5\% (5,000 individuals). 
Such a configuration is not very convenient in practical terms, since, on average, each selected domain will have fifteen observations. Also, under this particular scenario, the Horvitz-Thompson estimator becomes less efficient than both the calibration and composite ones. However, the proposed estimator has the property of high efficiency with a low $\overline{EERP}$ for all the simulation scenarios.\\

Another result in Table \ref{tabla_1_media} is that if the sample percentage increases $\left(f_{n}\uparrow\right)$, keing the number of domains fixed, then the $\overline{EERP}$ in all the estimators decreases. Since the sampling fraction inside each domain increases, therefore the variance of the estimates tends to decrease. On the other hand, if the number of domains increases $\left(f_{d}\uparrow\right)$ keing the sample percentage fixed, the $\overline{EERP}$ for all estimators increases. This is because, as the number of domains increases, the sample size per domain decreases and makes the estimates less efficient. The results obtained for the remaining scenarios are consistent with the ones in Table \ref{tabla_1_media} and they can be found in the Appendix.

\section{Application}\label{sec_application}

The information considered in this section corresponds to the one published in the OECD website \url{https://www.oecd.org/pisa/data/}. Plausible values for the PISA 2015 Mathematics Test are available in this dataset for each country and they will be used to estimate 
their average results. 
In addition, the auxiliary information that we consider to carry out
this application is composed of variables related to the learning context that, according to \citet*{trevino2016informe}, have a direct relationship with academic achievement. 
In particular, the auxiliary variables considered for each country in this case are: i) Gross Domestic Product (GDP), ii) Expenditure per student at secondary level (\% of GDP per capita), iii) Unemployment total (\% of total participation in the labour force as a national estimate), iv) Number of articles in scientific and technical publications, v) Expenditure on research and development (\% of GDP), vi) Public expenditure on education total (\% of GDP), vii) Gini index, viii) Percentage of schools with access to drinking water service, ix) Percentage of schools with access to electric service. 
The model that PISA uses to obtain the estimated ability of the students is shown below,
\begin{eqnarray}
\textsf{P}_{i}\left(\xi_{ik}=1\mid\theta_{k},\,a_{i},\,b_{i},\right) & = & \frac{e^{1.7a\left(\theta_{k}-b_{i}\right)}}{1+e^{1.7a\left(\theta_{k}-b_{i}\right)}}\,.\label{eq:modelo 2pl_aplicacion}
\end{eqnarray}
Also, the significant auxiliary variables and their associated coefficients, $\beta$, at a 10\% significance level are shown in Table \ref{tab:-estimados-y_pvalores}.

\begin{table}[H]
\noindent \begin{centering}
{\footnotesize{}}%
\begin{tabular}{lcc}
\hline 
{\footnotesize{}Covariate} & {\footnotesize{}$\hat{\beta}$} & {\footnotesize{}$p$-values}\tabularnewline
\hline 
{\footnotesize{}Intercept} & {\footnotesize{}$-5669.07$} & {\footnotesize{}$2.79\times10^{-5}$}\tabularnewline
{\footnotesize{}Public spending on education, total (\% of GDP)} & {\footnotesize{}$14.52$} & {\footnotesize{}$9.7310^{-3}$}\tabularnewline
{\footnotesize{}\% of schools with access to electricity service} & {\footnotesize{}$54.60$} & {\footnotesize{}$8.53\times10^{-5}$}\tabularnewline
{\footnotesize{}\% of schools with access to drinking water service
} & {\footnotesize{}$6.26$} & {\footnotesize{}$5.16\times10^{-2}$}\tabularnewline
{\footnotesize{}Research and development expenditure (\% of GDP)} & {\footnotesize{}$10.41$} & {\footnotesize{}$2.91\times10^{-2}$}\tabularnewline
{\footnotesize{}Unemployment, total (\% of total labor force participation)
} & {\footnotesize{}$-1.50$} & {\footnotesize{}$9.6\times10^{-2}$}\tabularnewline
\hline 
\end{tabular}{\footnotesize\par}
\par\end{centering}
\caption{Coefficients estimates, $\beta$, and their $p-$values \label{tab:-estimados-y_pvalores}}
\end{table}

In order to compute the estimated ability mean using the proposed estimator, we use directly Eq. \eqref{eq:estimador_propuesto}, and doing so, first, we estimate the variance of the random effect by means of Eq. \eqref{eq:est_beta_sigma_u}, obtaining that
$\hat{\sigma}_{u}=986.58$. The variances $\sigma_{d}^{2}$
and the direct estimates $\hat{\gamma}_{d}$ are those reported by PISA 2015.
Thus, in order to calculate the $MSE$ of $\hat{\gamma}_{d}^{\textsf{P}}$,
we must compute $g_{1d}\left(\hat{\sigma}_{u}^{2}\right)$, $g_{2d}\left(\hat{\sigma}_{u}^{2}\right)$, and $g_{3d}\left(\hat{\sigma}_{u}^{2}\right)$ as shown in Section \ref{sec_MSE}. The corresponding $MSE$ values of the proposed estimator $\hat{\gamma}_{d}^{\textsf{P}}$ are shown in Table \ref{tab:-MSE_est_pro}.

\noindent 
\begin{center}
{\scriptsize{}}%
\begin{tabular}{lcccccc}
\hline 
{\scriptsize{}Countries} & {\scriptsize{}$\sigma_{d}^{2}$} & {\scriptsize{}$B_{d}$} & {\scriptsize{}$1-B_{d}$} & {\scriptsize{}$\boldsymbol{x}_{d\,\textsf{A}}^{\textsf{T}}\hat{\boldsymbol\beta}$ } & {\scriptsize{}$\hat{\gamma}_{d}$} & {\scriptsize{}$\hat{\gamma}_{d}^{\textsf{P}}$}\tabularnewline
\hline 
{\scriptsize{}Albania } & {\scriptsize{}11.90 } & {\scriptsize{}0.01 } & {\scriptsize{}0.99 } & {\scriptsize{}430.00 } & {\scriptsize{}413.00 } & {\scriptsize{}413.00 }\tabularnewline
{\scriptsize{}Germany} & {\scriptsize{}8.35 } & {\scriptsize{}0.01 } & {\scriptsize{}0.99 } & {\scriptsize{}510.00 } & {\scriptsize{}506.00 } & {\scriptsize{}506.00 }\tabularnewline
{\scriptsize{}Australia } & {\scriptsize{}2.59 } & {\scriptsize{}0.00 } & {\scriptsize{}1.00 } & {\scriptsize{}476.00 } & {\scriptsize{}494.00 } & {\scriptsize{}494.00 }\tabularnewline
{\scriptsize{}Austria } & {\scriptsize{}8.18 } & {\scriptsize{}0.01 } & {\scriptsize{}0.99 } & {\scriptsize{}519.00 } & {\scriptsize{}497.00 } & {\scriptsize{}497.00 }\tabularnewline
{\scriptsize{}Belgium} & {\scriptsize{}5.52 } & {\scriptsize{}0.01 } & {\scriptsize{}0.99 } & {\scriptsize{}496.00 } & {\scriptsize{}507.00 } & {\scriptsize{}507.00 }\tabularnewline
{\scriptsize{}Brazil} & {\scriptsize{}8.18 } & {\scriptsize{}0.01 } & {\scriptsize{}0.99 } & {\scriptsize{}417.00 } & {\scriptsize{}377.00 } & {\scriptsize{}377.00 }\tabularnewline
{\scriptsize{}Bulgaria } & {\scriptsize{}15.60 } & {\scriptsize{}0.02 } & {\scriptsize{}0.98 } & {\scriptsize{}458.00 } & {\scriptsize{}441.00 } & {\scriptsize{}441.00 }\tabularnewline
{\scriptsize{}Canada} & {\scriptsize{}5.34 } & {\scriptsize{}0.01 } & {\scriptsize{}0.99 } & {\scriptsize{}502.00 } & {\scriptsize{}516.00 } & {\scriptsize{}516.00 }\tabularnewline
{\scriptsize{}Qatar } & {\scriptsize{}1.61 } & {\scriptsize{}0.00 } & {\scriptsize{}1.00 } & {\scriptsize{}472.00 } & {\scriptsize{}402.00 } & {\scriptsize{}402.00 }\tabularnewline
{\scriptsize{}Chile } & {\scriptsize{}6.45 } & {\scriptsize{}0.01 } & {\scriptsize{}0.99 } & {\scriptsize{}464.00 } & {\scriptsize{}423.00 } & {\scriptsize{}423.00 }\tabularnewline
{\scriptsize{}Colombia } & {\scriptsize{}5.24 } & {\scriptsize{}0.01 } & {\scriptsize{}0.99 } & {\scriptsize{}359.00 } & {\scriptsize{}390.00 } & {\scriptsize{}390.00 }\tabularnewline
{\scriptsize{}Korea} & {\scriptsize{}13.76 } & {\scriptsize{}0.01 } & {\scriptsize{}0.99 } & {\scriptsize{}529.00 } & {\scriptsize{}524.00 } & {\scriptsize{}524.00 }\tabularnewline
\textcolor{black}{\scriptsize{}Costa Rica } & {\scriptsize{}6.10 } & {\scriptsize{}0.01 } & {\scriptsize{}0.99 } & {\scriptsize{}416.00 } & {\scriptsize{}400.00 } & {\scriptsize{}400.00 }\tabularnewline
{\scriptsize{}Croatia} & {\scriptsize{}7.67 } & {\scriptsize{}0.01 } & {\scriptsize{}0.99 } & {\scriptsize{}465.00 } & {\scriptsize{}464.00 } & {\scriptsize{}464.00 }\tabularnewline
{\scriptsize{}Denmark} & {\scriptsize{}4.71 } & {\scriptsize{}0.00 } & {\scriptsize{}1.00 } & {\scriptsize{}513.00 } & {\scriptsize{}511.00 } & {\scriptsize{}511.00 }\tabularnewline
{\scriptsize{}Arab Emirates } & {\scriptsize{}5.81 } & {\scriptsize{}0.01 } & {\scriptsize{}0.99 } & {\scriptsize{}461.00 } & {\scriptsize{}427.00 } & {\scriptsize{}427.00 }\tabularnewline
{\scriptsize{}Slovakia} & {\scriptsize{}7.08 } & {\scriptsize{}0.01 } & {\scriptsize{}0.99 } & {\scriptsize{}478.00 } & {\scriptsize{}475.00 } & {\scriptsize{}475.00 }\tabularnewline
{\scriptsize{}Slovenia} & {\scriptsize{}1.59 } & {\scriptsize{}0.00 } & {\scriptsize{}1.00 } & {\scriptsize{}495.00 } & {\scriptsize{}510.00 } & {\scriptsize{}510.00 }\tabularnewline
{\scriptsize{}Spain} & {\scriptsize{}4.62 } & {\scriptsize{}0.00 } & {\scriptsize{}1.00 } & {\scriptsize{}458.00 } & {\scriptsize{}486.00 } & {\scriptsize{}486.00 }\tabularnewline
{\scriptsize{}United States} & {\scriptsize{}10.05 } & {\scriptsize{}0.01 } & {\scriptsize{}0.99 } & {\scriptsize{}505.00 } & {\scriptsize{}470.00 } & {\scriptsize{}470.00 }\tabularnewline
{\scriptsize{}Estonia } & {\scriptsize{}4.16 } & {\scriptsize{}0.00 } & {\scriptsize{}1.00 } & {\scriptsize{}497.00 } & {\scriptsize{}520.00 } & {\scriptsize{}520.00 }\tabularnewline
{\scriptsize{}Russian Federation } & {\scriptsize{}9.67 } & {\scriptsize{}0.01 } & {\scriptsize{}0.99 } & {\scriptsize{}456.00 } & {\scriptsize{}494.00 } & {\scriptsize{}494.00 }\tabularnewline
{\scriptsize{}Finland} & {\scriptsize{}5.34 } & {\scriptsize{}0.01 } & {\scriptsize{}0.99 } & {\scriptsize{}507.00 } & {\scriptsize{}511.00 } & {\scriptsize{}511.00 }\tabularnewline
{\scriptsize{}France} & {\scriptsize{}4.41 } & {\scriptsize{}0.00 } & {\scriptsize{}1.00 } & {\scriptsize{}470.00 } & {\scriptsize{}493.00 } & {\scriptsize{}493.00 }\tabularnewline
{\scriptsize{}Greece} & {\scriptsize{}14.06 } & {\scriptsize{}0.01 } & {\scriptsize{}0.99 } & {\scriptsize{}447.00 } & {\scriptsize{}454.00 } & {\scriptsize{}454.00 }\tabularnewline
{\scriptsize{}Hong Kong-China } & {\scriptsize{}8.88 } & {\scriptsize{}0.01 } & {\scriptsize{}0.99 } & {\scriptsize{}505.00 } & {\scriptsize{}548.00 } & {\scriptsize{}548.00 }\tabularnewline
{\scriptsize{}Hungary} & {\scriptsize{}6.40 } & {\scriptsize{}0.01 } & {\scriptsize{}0.99 } & {\scriptsize{}487.00 } & {\scriptsize{}477.00 } & {\scriptsize{}477.00 }\tabularnewline
{\scriptsize{}Indonesia } & {\scriptsize{}9.49 } & {\scriptsize{}0.01 } & {\scriptsize{}0.99 } & {\scriptsize{}420.00 } & {\scriptsize{}386.00 } & {\scriptsize{}386.00 }\tabularnewline
{\scriptsize{}Ireland} & {\scriptsize{}4.20 } & {\scriptsize{}0.00 } & {\scriptsize{}1.00 } & {\scriptsize{}453.00 } & {\scriptsize{}504.00 } & {\scriptsize{}504.00 }\tabularnewline
{\scriptsize{}Iceland} & {\scriptsize{}3.96 } & {\scriptsize{}0.00 } & {\scriptsize{}1.00 } & {\scriptsize{}542.00 } & {\scriptsize{}488.00 } & {\scriptsize{}488.00 }\tabularnewline
{\scriptsize{}Israel } & {\scriptsize{}13.18 } & {\scriptsize{}0.01 } & {\scriptsize{}0.99 } & {\scriptsize{}512.00 } & {\scriptsize{}470.00 } & {\scriptsize{}471.00 }\tabularnewline
{\scriptsize{}Italy} & {\scriptsize{}8.12 } & {\scriptsize{}0.01 } & {\scriptsize{}0.99 } & {\scriptsize{}469.00 } & {\scriptsize{}490.00 } & {\scriptsize{}490.00 }\tabularnewline
{\scriptsize{}Japan} & {\scriptsize{}9.00 } & {\scriptsize{}0.01 } & {\scriptsize{}0.99 } & {\scriptsize{}519.00 } & {\scriptsize{}532.00 } & {\scriptsize{}532.00 }\tabularnewline
{\scriptsize{}Jordan } & {\scriptsize{}7.02 } & {\scriptsize{}0.01 } & {\scriptsize{}0.99 } & {\scriptsize{}420.00 } & {\scriptsize{}380.00 } & {\scriptsize{}380.00 }\tabularnewline
{\scriptsize{}Latvia} & {\scriptsize{}3.50 } & {\scriptsize{}0.00 } & {\scriptsize{}1.00 } & {\scriptsize{}461.00 } & {\scriptsize{}482.00 } & {\scriptsize{}482.00 }\tabularnewline
{\scriptsize{}Lithuania} & {\scriptsize{}5.43 } & {\scriptsize{}0.01 } & {\scriptsize{}0.99 } & {\scriptsize{}454.00 } & {\scriptsize{}478.00 } & {\scriptsize{}478.00 }\tabularnewline
{\scriptsize{}Luxembourg} & {\scriptsize{}1.61 } & {\scriptsize{}0.00 } & {\scriptsize{}1.00 } & {\scriptsize{}477.00 } & {\scriptsize{}486.00 } & {\scriptsize{}486.00 }\tabularnewline
{\scriptsize{}Macao-China } & {\scriptsize{}1.23 } & {\scriptsize{}0.00 } & {\scriptsize{}1.00 } & {\scriptsize{}491.00 } & {\scriptsize{}544.00 } & {\scriptsize{}544.00 }\tabularnewline
{\scriptsize{}Mexico } & {\scriptsize{}5.02 } & {\scriptsize{}0.01 } & {\scriptsize{}0.99 } & {\scriptsize{}398.00 } & {\scriptsize{}408.00 } & {\scriptsize{}408.00 }\tabularnewline
\end{tabular}{\scriptsize\par}
\par\end{center}

\noindent \begin{center}
{\scriptsize{}}%
\begin{tabular}{lcccccc}
{\scriptsize{}Montenegro } & {\scriptsize{}2.13 } & {\scriptsize{}0.00 } & {\scriptsize{}1.00 } & {\scriptsize{}437.00 } & {\scriptsize{}418.00 } & {\scriptsize{}418.00 }\tabularnewline
{\scriptsize{}Norway} & {\scriptsize{}4.97 } & {\scriptsize{}0.01 } & {\scriptsize{}0.99 } & {\scriptsize{}482.00 } & {\scriptsize{}502.00 } & {\scriptsize{}502.00 }\tabularnewline
{\scriptsize{}New Zealand } & {\scriptsize{}5.15 } & {\scriptsize{}0.01 } & {\scriptsize{}0.99 } & {\scriptsize{}470.00 } & {\scriptsize{}495.00 } & {\scriptsize{}495.00 }\tabularnewline
{\scriptsize{}Netherlands} & {\scriptsize{}4.88 } & {\scriptsize{}0.00 } & {\scriptsize{}1.00 } & {\scriptsize{}504.00 } & {\scriptsize{}512.00 } & {\scriptsize{}512.00 }\tabularnewline
{\scriptsize{}Peru} & {\scriptsize{}7.34 } & {\scriptsize{}0.01 } & {\scriptsize{}0.99 } & {\scriptsize{}424.00 } & {\scriptsize{}387.00 } & {\scriptsize{}387.00 }\tabularnewline
{\scriptsize{}Poland} & {\scriptsize{}5.71 } & {\scriptsize{}0.01 } & {\scriptsize{}0.99 } & {\scriptsize{}480.00 } & {\scriptsize{}504.00 } & {\scriptsize{}504.00 }\tabularnewline
{\scriptsize{}Portugal } & {\scriptsize{}6.20 } & {\scriptsize{}0.01 } & {\scriptsize{}0.99 } & {\scriptsize{}482.00 } & {\scriptsize{}492.00 } & {\scriptsize{}492.00 }\tabularnewline
{\scriptsize{}Czech Republic } & {\scriptsize{}5.76 } & {\scriptsize{}0.01 } & {\scriptsize{}0.99 } & {\scriptsize{}513.00 } & {\scriptsize{}492.00 } & {\scriptsize{}492.00 }\tabularnewline
{\scriptsize{}Romania} & {\scriptsize{}14.36 } & {\scriptsize{}0.01 } & {\scriptsize{}0.99 } & {\scriptsize{}457.00 } & {\scriptsize{}444.00 } & {\scriptsize{}444.00 }\tabularnewline
{\scriptsize{}Singapore} & {\scriptsize{}2.16 } & {\scriptsize{}0.00 } & {\scriptsize{}1.00 } & {\scriptsize{}511.00 } & {\scriptsize{}564.00 } & {\scriptsize{}564.00 }\tabularnewline
{\scriptsize{}Sweden} & {\scriptsize{}10.05 } & {\scriptsize{}0.01 } & {\scriptsize{}0.99 } & {\scriptsize{}491.00 } & {\scriptsize{}494.00 } & {\scriptsize{}494.00 }\tabularnewline
{\scriptsize{}Switzerland} & {\scriptsize{}8.53 } & {\scriptsize{}0.01 } & {\scriptsize{}0.99 } & {\scriptsize{}519.00 } & {\scriptsize{}521.00 } & {\scriptsize{}521.00 }\tabularnewline
{\scriptsize{}Thailand} & {\scriptsize{}9.18 } & {\scriptsize{}0.01 } & {\scriptsize{}0.99 } & {\scriptsize{}456.00 } & {\scriptsize{}415.00 } & {\scriptsize{}415.00 }\tabularnewline
{\scriptsize{}Tunisia} & {\scriptsize{}8.70 } & {\scriptsize{}0.01 } & {\scriptsize{}0.99 } & {\scriptsize{}355.00 } & {\scriptsize{}367.00 } & {\scriptsize{}367.00 }\tabularnewline
{\scriptsize{}Turkey} & {\scriptsize{}17.06 } & {\scriptsize{}0.02 } & {\scriptsize{}0.98 } & {\scriptsize{}473.00 } & {\scriptsize{}420.00 } & {\scriptsize{}421.00 }\tabularnewline
{\scriptsize{}Vietnam } & {\scriptsize{}19.89 } & {\scriptsize{}0.02 } & {\scriptsize{}0.98 } & {\scriptsize{}439.00 } & {\scriptsize{}495.00 } & {\scriptsize{}494.00 }\tabularnewline
\hline 
\end{tabular}{\scriptsize\par}
\par\end{center}

\begin{table}[H]
\caption{Estimate of the estimator $\hat{\gamma}_{d}^{\textsf{P}}$ by country. \label{tab:Estimaci=0000F3n-por-pa=0000EDs_estimador_propuesto}}
\end{table}

\noindent \begin{center}
{\footnotesize{}}%
\begin{tabular}{lcccc}
\hline 
{\scriptsize{}Countries} & {\footnotesize{}$g_{1d}\left(\hat{\sigma}_{u}^{2}\right)$ } & {\footnotesize{}$g_{2d}\left(\hat{\sigma}_{u}^{2}\right)$} & {\footnotesize{}$g_{3d}\left(\hat{\sigma}_{u}^{2}\right)$} & {\footnotesize{}$MSE\left(\hat{\gamma}_{d}^{\textsf{P}}\right)$ }\tabularnewline
\hline 
{\scriptsize{}Albania } & {\footnotesize{}11.7606 } & {\footnotesize{}0.0149 } & {\footnotesize{}0.0051 } & {\footnotesize{}11.7810 }\tabularnewline
{\scriptsize{}Germany} & {\footnotesize{}8.2820 } & {\footnotesize{}0.0038 } & {\footnotesize{}0.0025 } & {\footnotesize{}8.2880 }\tabularnewline
{\scriptsize{}Australia } & {\footnotesize{}2.5853 } & {\footnotesize{}0.0006 } & {\footnotesize{}0.0002 } & {\footnotesize{}2.5860 }\tabularnewline
{\scriptsize{}Austria } & {\footnotesize{}8.1123 } & {\footnotesize{}0.0048 } & {\footnotesize{}0.0024 } & {\footnotesize{}8.1200 }\tabularnewline
{\scriptsize{}Belgium} & {\footnotesize{}5.4918 } & {\footnotesize{}0.0011 } & {\footnotesize{}0.0011 } & {\footnotesize{}5.4940 }\tabularnewline
{\scriptsize{}Brazil} & {\footnotesize{}8.1123 } & {\footnotesize{}0.0096 } & {\footnotesize{}0.0024 } & {\footnotesize{}8.1240 }\tabularnewline
{\scriptsize{}Bulgaria } & {\footnotesize{}15.3596 } & {\footnotesize{}0.0167 } & {\footnotesize{}0.0087 } & {\footnotesize{}15.3850 }\tabularnewline
{\scriptsize{}Canada} & {\footnotesize{}5.3074 } & {\footnotesize{}0.0013 } & {\footnotesize{}0.0010 } & {\footnotesize{}5.3100 }\tabularnewline
{\scriptsize{}Qatar } & {\footnotesize{}1.6103 } & {\footnotesize{}0.0004 } & {\footnotesize{}0.0001 } & {\footnotesize{}1.6110 }\tabularnewline
{\scriptsize{}Chile } & {\footnotesize{}6.4097 } & {\footnotesize{}0.0031 } & {\footnotesize{}0.0015 } & {\footnotesize{}6.4140 }\tabularnewline
{\scriptsize{}Colombia } & {\footnotesize{}5.2164 } & {\footnotesize{}0.0145 } & {\footnotesize{}0.0010 } & {\footnotesize{}5.2320 }\tabularnewline
{\scriptsize{}Korea} & {\footnotesize{}13.5747 } & {\footnotesize{}0.0281 } & {\footnotesize{}0.0068 } & {\footnotesize{}13.6100 }\tabularnewline
\textcolor{black}{\scriptsize{}Costa Rica } & {\footnotesize{}6.0634 } & {\footnotesize{}0.0045 } & {\footnotesize{}0.0014 } & {\footnotesize{}6.0690 }\tabularnewline
{\scriptsize{}Croatia} & {\footnotesize{}7.6137 } & {\footnotesize{}0.0050 } & {\footnotesize{}0.0022 } & {\footnotesize{}7.6210 }\tabularnewline
{\scriptsize{}Denmark} & {\footnotesize{}4.6865 } & {\footnotesize{}0.0013 } & {\footnotesize{}0.0008 } & {\footnotesize{}4.6890 }\tabularnewline
{\scriptsize{}Arab Emirates } & {\footnotesize{}5.7741 } & {\footnotesize{}0.0016 } & {\footnotesize{}0.0012 } & {\footnotesize{}5.7770 }\tabularnewline
{\scriptsize{}Slovakia} & {\footnotesize{}7.0252 } & {\footnotesize{}0.0020 } & {\footnotesize{}0.0018 } & {\footnotesize{}7.0290 }\tabularnewline
{\scriptsize{}Slovenia} & {\footnotesize{}1.5850 } & {\footnotesize{}0.0001 } & {\footnotesize{}0.0001 } & {\footnotesize{}1.5850 }\tabularnewline
{\scriptsize{}Spain} & {\footnotesize{}4.6009 } & {\footnotesize{}0.0040 } & {\footnotesize{}0.0008 } & {\footnotesize{}4.6060 }\tabularnewline
{\scriptsize{}United States} & {\footnotesize{}9.9476 } & {\footnotesize{}0.0050 } & {\footnotesize{}0.0037 } & {\footnotesize{}9.9560 }\tabularnewline
{\scriptsize{}Estonia } & {\footnotesize{}4.1441 } & {\footnotesize{}0.0007 } & {\footnotesize{}0.0006 } & {\footnotesize{}4.1450 }\tabularnewline
{\scriptsize{}Russian Federation } & {\footnotesize{}9.5782 } & {\footnotesize{}0.0077 } & {\footnotesize{}0.0034 } & {\footnotesize{}9.5890 }\tabularnewline
{\scriptsize{}Finland} & {\footnotesize{}5.3074 } & {\footnotesize{}0.0018 } & {\footnotesize{}0.0010 } & {\footnotesize{}5.3100 }\tabularnewline
{\scriptsize{}France} & {\footnotesize{}4.3904 } & {\footnotesize{}0.0021 } & {\footnotesize{}0.0007 } & {\footnotesize{}4.3930 }\tabularnewline
{\scriptsize{}Greece} & {\footnotesize{}13.8649 } & {\footnotesize{}0.0496 } & {\footnotesize{}0.0071 } & {\footnotesize{}13.9220 }\tabularnewline
{\scriptsize{}Hong Kong-China } & {\footnotesize{}8.8012 } & {\footnotesize{}0.0101 } & {\footnotesize{}0.0029 } & {\footnotesize{}8.8140 }\tabularnewline
{\scriptsize{}Hungary} & {\footnotesize{}6.3596 } & {\footnotesize{}0.0013 } & {\footnotesize{}0.0015 } & {\footnotesize{}6.3620 }\tabularnewline
{\scriptsize{}Indonesia } & {\footnotesize{}9.3961 } & {\footnotesize{}0.0079 } & {\footnotesize{}0.0033 } & {\footnotesize{}9.4070 }\tabularnewline
{\scriptsize{}Ireland} & {\footnotesize{}4.1847 } & {\footnotesize{}0.0010 } & {\footnotesize{}0.0007 } & {\footnotesize{}4.1860 }\tabularnewline
{\scriptsize{}Iceland} & {\footnotesize{}3.9443 } & {\footnotesize{}0.0045 } & {\footnotesize{}0.0006 } & {\footnotesize{}3.9490 }\tabularnewline
{\scriptsize{}Israel } & {\footnotesize{}13.0032 } & {\footnotesize{}0.0311 } & {\footnotesize{}0.0062 } & {\footnotesize{}13.0410 }\tabularnewline
{\scriptsize{}Italy} & {\footnotesize{}8.0562 } & {\footnotesize{}0.0024 } & {\footnotesize{}0.0024 } & {\footnotesize{}8.0610 }\tabularnewline
{\scriptsize{}Japan} & {\footnotesize{}8.9186 } & {\footnotesize{}0.0063 } & {\footnotesize{}0.0029 } & {\footnotesize{}8.9280 }\tabularnewline
{\scriptsize{}Jordan } & {\footnotesize{}6.9729 } & {\footnotesize{}0.0048 } & {\footnotesize{}0.0018 } & {\footnotesize{}6.9790 }\tabularnewline
{\scriptsize{}Latvia} & {\footnotesize{}3.4845 } & {\footnotesize{}0.0005 } & {\footnotesize{}0.0005 } & {\footnotesize{}3.4850 }\tabularnewline
{\scriptsize{}Lithuania} & {\footnotesize{}5.3992 } & {\footnotesize{}0.0023 } & {\footnotesize{}0.0011 } & {\footnotesize{}5.4030 }\tabularnewline
{\scriptsize{}Luxembourg} & {\footnotesize{}1.6103 } & {\footnotesize{}0.0001 } & {\footnotesize{}0.0001 } & {\footnotesize{}1.6100 }\tabularnewline
{\scriptsize{}Macao-China } & {\footnotesize{}1.2306 } & {\footnotesize{}0.0002 } & {\footnotesize{}0.0001 } & {\footnotesize{}1.2310 }\tabularnewline
{\scriptsize{}Mexico } & {\footnotesize{}4.9922 } & {\footnotesize{}0.0048 } & {\footnotesize{}0.0009 } & 
{\footnotesize{}4.9980 }\tabularnewline
\end{tabular}{\footnotesize\par}
\par\end{center}

\noindent \begin{center}
{\footnotesize{}}%
\begin{tabular}{lcccc}
{\scriptsize{}Montenegro } & {\footnotesize{}2.1270 } & {\footnotesize{}0.0005 } & {\footnotesize{}0.0002 } & {\footnotesize{}2.1280 }\tabularnewline
{\scriptsize{}Norway} & {\footnotesize{}4.9480 } & {\footnotesize{}0.0020 } & {\footnotesize{}0.0009 } & {\footnotesize{}4.9510 }\tabularnewline
{\scriptsize{}New Zealand } & {\footnotesize{}5.1261 } & {\footnotesize{}0.0025 } & {\footnotesize{}0.0010 } & {\footnotesize{}5.1300 }\tabularnewline
{\scriptsize{}Netherlands} & {\footnotesize{}4.8600 } & {\footnotesize{}0.0010 } & {\footnotesize{}0.0009 } & {\footnotesize{}4.8620 }\tabularnewline
{\scriptsize{}Peru} & {\footnotesize{}7.2898 } & {\footnotesize{}0.0083 } & {\footnotesize{}0.0020 } & {\footnotesize{}7.3000 }\tabularnewline
{\scriptsize{}Poland} & {\footnotesize{}5.6792 } & {\footnotesize{}0.0011 } & {\footnotesize{}0.0012 } & {\footnotesize{}5.6820 }\tabularnewline
{\scriptsize{}Portugal } & {\footnotesize{}6.1614 } & {\footnotesize{}0.0021 } & {\footnotesize{}0.0014 } & {\footnotesize{}6.1650 }\tabularnewline
{\scriptsize{}Czech Republic } & {\footnotesize{}5.7266 } & {\footnotesize{}0.0023 } & {\footnotesize{}0.0012 } & {\footnotesize{}5.7300 }\tabularnewline
{\scriptsize{}Romania} & {\footnotesize{}14.1580 } & {\footnotesize{}0.0259 } & {\footnotesize{}0.0074 } & {\footnotesize{}14.1910 }\tabularnewline
{\scriptsize{}Singapore} & {\footnotesize{}2.1562 } & {\footnotesize{}0.0002 } & {\footnotesize{}0.0002 } & {\footnotesize{}2.1570 }\tabularnewline
{\scriptsize{}Sweden} & {\footnotesize{}9.9476 } & {\footnotesize{}0.0123 } & {\footnotesize{}0.0037 } & {\footnotesize{}9.9640 }\tabularnewline
{\scriptsize{}Switzerland} & {\footnotesize{}8.4533 } & {\footnotesize{}0.0058 } & {\footnotesize{}0.0026 } & {\footnotesize{}8.4620 }\tabularnewline
{\scriptsize{}Thailand} & {\footnotesize{}9.0963 } & {\footnotesize{}0.0093 } & {\footnotesize{}0.0031 } & {\footnotesize{}9.1090 }\tabularnewline
{\scriptsize{}Tunisia} & {\footnotesize{}8.6264 } & {\footnotesize{}0.0234 } & {\footnotesize{}0.0028 } & {\footnotesize{}8.6530 }\tabularnewline
{\scriptsize{}Turkey} & {\footnotesize{}16.7670 } & {\footnotesize{}0.0109 } & {\footnotesize{}0.0103 } & {\footnotesize{}16.7880 }\tabularnewline
{\scriptsize{}Vietnam } & {\footnotesize{}19.4985 } & {\footnotesize{}0.1504 } & {\footnotesize{}0.0139 } & {\footnotesize{}19.6630 }\tabularnewline
\hline 
\end{tabular}{\footnotesize\par}
\par\end{center}

\begin{table}[H]
\caption{$MSE$ of $\hat{\gamma}_{d}^{\textsf{P}}$ by country.\label{tab:-MSE_est_pro}}
\end{table}

\noindent \begin{center}
{\scriptsize{}}%
\begin{tabular}{lccccc}
\hline 
{\scriptsize{}Countries} & {\scriptsize{}$\hat{\gamma}_{d}$} & {\scriptsize{}$CVE\left(100\%\right)$} & {\scriptsize{}$\hat{\gamma}_{d}^{\textsf{P}}$} & {\scriptsize{}$EER_{d}\left(100\%\right)$} & {\scriptsize{}$Dif_{rel}\,\left(100\%\right)$ }\tabularnewline
\hline 
{\scriptsize{}Albania } & {\scriptsize{}413.0000 } & {\scriptsize{}0.8354 } & {\scriptsize{}413.0000 } & {\scriptsize{}0.8308 } & {\scriptsize{}0.9810 }\tabularnewline
{\scriptsize{}Germany} & {\scriptsize{}506.0000 } & {\scriptsize{}0.5711 } & {\scriptsize{}506.0000 } & {\scriptsize{}0.5690 } & {\scriptsize{}0.7332 }\tabularnewline
{\scriptsize{}Australia } & {\scriptsize{}494.0000 } & {\scriptsize{}0.3259 } & {\scriptsize{}494.0000 } & {\scriptsize{}0.3256 } & {\scriptsize{}0.2202 }\tabularnewline
{\scriptsize{}Austria } & {\scriptsize{}497.0000 } & {\scriptsize{}0.5755 } & {\scriptsize{}497.0000 } & {\scriptsize{}0.5732 } & {\scriptsize{}0.7043 }\tabularnewline
{\scriptsize{}Belgium} & {\scriptsize{}507.0000 } & {\scriptsize{}0.4635 } & {\scriptsize{}507.0000 } & {\scriptsize{}0.4624 } & {\scriptsize{}0.4953 }\tabularnewline
{\scriptsize{}Brazil} & {\scriptsize{}377.0000 } & {\scriptsize{}0.7586 } & {\scriptsize{}377.0000 } & {\scriptsize{}0.7555 } & {\scriptsize{}0.6457 }\tabularnewline
{\scriptsize{}Bulgaria } & {\scriptsize{}441.0000 } & {\scriptsize{}0.8957 } & {\scriptsize{}441.0000 } & {\scriptsize{}0.8891 } & {\scriptsize{}1.3383 }\tabularnewline
{\scriptsize{}Canada} & {\scriptsize{}516.0000 } & {\scriptsize{}0.4477 } & {\scriptsize{}516.0000 } & {\scriptsize{}0.4467 } & {\scriptsize{}0.4747 }\tabularnewline
{\scriptsize{}Qatar } & {\scriptsize{}402.0000 } & {\scriptsize{}0.3159 } & {\scriptsize{}402.0000 } & {\scriptsize{}0.3156 } & {\scriptsize{}0.1256 }\tabularnewline
{\scriptsize{}Chile } & {\scriptsize{}423.0000 } & {\scriptsize{}0.6005 } & {\scriptsize{}423.0000 } & {\scriptsize{}0.5984 } & {\scriptsize{}0.5547 }\tabularnewline
{\scriptsize{}Colombia } & {\scriptsize{}390.0000 } & {\scriptsize{}0.5872 } & {\scriptsize{}390.0000 } & {\scriptsize{}0.5868 } & {\scriptsize{}0.2132 }\tabularnewline
{\scriptsize{}Korea} & {\scriptsize{}524.0000 } & {\scriptsize{}0.7080 } & {\scriptsize{}524.0000 } & {\scriptsize{}0.7041 } & {\scriptsize{}1.0728 }\tabularnewline
\textcolor{black}{\scriptsize{}Costa Rica } & {\scriptsize{}400.0000 } & {\scriptsize{}0.6175 } & {\scriptsize{}400.0000 } & {\scriptsize{}0.6158 } & {\scriptsize{}0.4957 }\tabularnewline
{\scriptsize{}Croatia} & {\scriptsize{}464.0000 } & {\scriptsize{}0.5970 } & {\scriptsize{}464.0000 } & {\scriptsize{}0.5950 } & {\scriptsize{}0.6502 }\tabularnewline
{\scriptsize{}Denmark} & {\scriptsize{}511.0000 } & {\scriptsize{}0.4247 } & {\scriptsize{}511.0000 } & {\scriptsize{}0.4238 } & {\scriptsize{}0.4118 }\tabularnewline
{\scriptsize{}Arab Emirates } & {\scriptsize{}427.0000 } & {\scriptsize{}0.5644 } & {\scriptsize{}427.0000 } & {\scriptsize{}0.5627 } & {\scriptsize{}0.5146 }\tabularnewline
{\scriptsize{}Slovakia} & {\scriptsize{}475.0000 } & {\scriptsize{}0.5600 } & {\scriptsize{}475.0000 } & {\scriptsize{}0.5582 } & {\scriptsize{}0.6316 }\tabularnewline
{\scriptsize{}Slovenia} & {\scriptsize{}510.0000 } & {\scriptsize{}0.2471 } & {\scriptsize{}510.0000 } & {\scriptsize{}0.2469 } & {\scriptsize{}0.1436 }\tabularnewline
{\scriptsize{}Spain} & {\scriptsize{}486.0000 } & {\scriptsize{}0.4424 } & {\scriptsize{}486.0000 } & {\scriptsize{}0.4417 } & {\scriptsize{}0.3465 }\tabularnewline
{\scriptsize{}United States} & {\scriptsize{}470.0000 } & {\scriptsize{}0.6745 } & {\scriptsize{}470.0000 } & {\scriptsize{}0.6710 } & {\scriptsize{}0.8854 }\tabularnewline
{\scriptsize{}Estonia } & {\scriptsize{}520.0000 } & {\scriptsize{}0.3923 } & {\scriptsize{}520.0000 } & {\scriptsize{}0.3916 } & {\scriptsize{}0.3724 }\tabularnewline
{\scriptsize{}Russian Federation } & {\scriptsize{}494.0000 } & {\scriptsize{}0.6296 } & {\scriptsize{}494.0000 } & {\scriptsize{}0.6274 } & {\scriptsize{}0.8207 }\tabularnewline
{\scriptsize{}Finland} & {\scriptsize{}511.0000 } & {\scriptsize{}0.4521 } & {\scriptsize{}511.0000 } & {\scriptsize{}0.4510 } & {\scriptsize{}0.4652 }\tabularnewline
{\scriptsize{}France} & {\scriptsize{}493.0000 } & {\scriptsize{}0.4260 } & {\scriptsize{}493.0000 } & {\scriptsize{}0.4253 } & {\scriptsize{}0.3645 }\tabularnewline
{\scriptsize{}Greece} & {\scriptsize{}454.0000 } & {\scriptsize{}0.8260 } & {\scriptsize{}454.0000 } & {\scriptsize{}0.8222 } & {\scriptsize{}0.9518 }\tabularnewline
{\scriptsize{}Hong Kong-China } & {\scriptsize{}548.0000 } & {\scriptsize{}0.5438 } & {\scriptsize{}548.0000 } & {\scriptsize{}0.5422 } & {\scriptsize{}0.7138 }\tabularnewline
{\scriptsize{}Hungary} & {\scriptsize{}477.0000 } & {\scriptsize{}0.5304 } & {\scriptsize{}477.0000 } & {\scriptsize{}0.5288 } & {\scriptsize{}0.5770 }\tabularnewline
{\scriptsize{}Indonesia } & {\scriptsize{}386.0000 } & {\scriptsize{}0.7979 } & {\scriptsize{}386.0000 } & {\scriptsize{}0.7941 } & {\scriptsize{}0.7996 }\tabularnewline
{\scriptsize{}Ireland} & {\scriptsize{}504.0000 } & {\scriptsize{}0.4067 } & {\scriptsize{}504.0000 } & {\scriptsize{}0.4062 } & {\scriptsize{}0.3698 }\tabularnewline
{\scriptsize{}Iceland} & {\scriptsize{}488.0000 } & {\scriptsize{}0.4078 } & {\scriptsize{}488.0000 } & {\scriptsize{}0.4071 } & {\scriptsize{}0.2573 }\tabularnewline
{\scriptsize{}Israel } & {\scriptsize{}470.0000 } & {\scriptsize{}0.7723 } & {\scriptsize{}471.0000 } & {\scriptsize{}0.7676 } & {\scriptsize{}0.9874 }\tabularnewline
{\scriptsize{}Italy} & {\scriptsize{}490.0000 } & {\scriptsize{}0.5816 } & {\scriptsize{}490.0000 } & {\scriptsize{}0.5797 } & {\scriptsize{}0.7283 }\tabularnewline
{\scriptsize{}Japan} & {\scriptsize{}532.0000 } & {\scriptsize{}0.5639 } & {\scriptsize{}532.0000 } & {\scriptsize{}0.5619 } & {\scriptsize{}0.7682 }\tabularnewline
{\scriptsize{}Jordan } & {\scriptsize{}380.0000 } & {\scriptsize{}0.6974 } & {\scriptsize{}380.0000 } & {\scriptsize{}0.6948 } & {\scriptsize{}0.5869 }\tabularnewline
{\scriptsize{}Latvia} & {\scriptsize{}482.0000 } & {\scriptsize{}0.3880 } & {\scriptsize{}482.0000 } & {\scriptsize{}0.3874 } & {\scriptsize{}0.3140 }\tabularnewline
{\scriptsize{}Lithuania} & {\scriptsize{}478.0000 } & {\scriptsize{}0.4874 } & {\scriptsize{}478.0000 } & {\scriptsize{}0.4864 } & {\scriptsize{}0.4650 }\tabularnewline
{\scriptsize{}Luxembourg} & {\scriptsize{}486.0000 } & {\scriptsize{}0.2613 } & {\scriptsize{}486.0000 } & {\scriptsize{}0.2611 } & {\scriptsize{}0.1439 }\tabularnewline
{\scriptsize{}Macao-China } & {\scriptsize{}544.0000 } & {\scriptsize{}0.2040 } & {\scriptsize{}544.0000 } & {\scriptsize{}0.2040 } & {\scriptsize{}0.0966 }\tabularnewline
{\scriptsize{}Mexico } & {\scriptsize{}408.0000 } & {\scriptsize{}0.5490 } & {\scriptsize{}408.0000 } & {\scriptsize{}0.5481 } & {\scriptsize{}0.3736 }\tabularnewline
\end{tabular}{\scriptsize\par}
\par\end{center}

\noindent \begin{center}
{\scriptsize{}}%
\begin{tabular}{lccccc}
{\scriptsize{}Montenegro } & {\scriptsize{}418.0000 } & {\scriptsize{}0.3493 } & {\scriptsize{}418.0000 } & {\scriptsize{}0.3489 } & {\scriptsize{}0.1740 }\tabularnewline
{\scriptsize{}Norway} & {\scriptsize{}502.0000 } & {\scriptsize{}0.4442 } & {\scriptsize{}502.0000 } & {\scriptsize{}0.4434 } & {\scriptsize{}0.4253 }\tabularnewline
{\scriptsize{}New Zealand } & {\scriptsize{}495.0000 } & {\scriptsize{}0.4586 } & {\scriptsize{}495.0000 } & {\scriptsize{}0.4577 } & {\scriptsize{}0.4341 }\tabularnewline
{\scriptsize{}Netherlands} & {\scriptsize{}512.0000 } & {\scriptsize{}0.4316 } & {\scriptsize{}512.0000 } & {\scriptsize{}0.4307 } & {\scriptsize{}0.4370 }\tabularnewline
{\scriptsize{}Peru} & {\scriptsize{}387.0000 } & {\scriptsize{}0.7003 } & {\scriptsize{}387.0000 } & {\scriptsize{}0.6978 } & {\scriptsize{}0.5727 }\tabularnewline
{\scriptsize{}Poland} & {\scriptsize{}504.0000 } & {\scriptsize{}0.4742 } & {\scriptsize{}504.0000 } & {\scriptsize{}0.4731 } & {\scriptsize{}0.5140 }\tabularnewline
{\scriptsize{}Portugal } & {\scriptsize{}492.0000 } & {\scriptsize{}0.5061 } & {\scriptsize{}492.0000 } & {\scriptsize{}0.5048 } & {\scriptsize{}0.5459 }\tabularnewline
{\scriptsize{}Czech Republic } & {\scriptsize{}492.0000 } & {\scriptsize{}0.4878 } & {\scriptsize{}492.0000 } & {\scriptsize{}0.4865 } & {\scriptsize{}0.4978 }\tabularnewline
{\scriptsize{}Romania} & {\scriptsize{}444.0000 } & {\scriptsize{}0.8536 } & {\scriptsize{}444.0000 } & {\scriptsize{}0.8483 } & {\scriptsize{}1.1518 }\tabularnewline
{\scriptsize{}Singapore} & {\scriptsize{}564.0000 } & {\scriptsize{}0.2606 } & {\scriptsize{}564.0000 } & {\scriptsize{}0.2604 } & {\scriptsize{}0.1922 }\tabularnewline
{\scriptsize{}Sweden} & {\scriptsize{}494.0000 } & {\scriptsize{}0.6417 } & {\scriptsize{}494.0000 } & {\scriptsize{}0.6391 } & {\scriptsize{}0.8131 }\tabularnewline
{\scriptsize{}Switzerland} & {\scriptsize{}521.0000 } & {\scriptsize{}0.5605 } & {\scriptsize{}521.0000 } & {\scriptsize{}0.5584 } & {\scriptsize{}0.7269 }\tabularnewline
{\scriptsize{}Thailand} & {\scriptsize{}415.0000 } & {\scriptsize{}0.7301 } & {\scriptsize{}415.0000 } & {\scriptsize{}0.7267 } & {\scriptsize{}0.7536 }\tabularnewline
{\scriptsize{}Tunisia} & {\scriptsize{}367.0000 } & {\scriptsize{}0.8038 } & {\scriptsize{}367.0000 } & {\scriptsize{}0.8019 } & {\scriptsize{}0.5419 }\tabularnewline
{\scriptsize{}Turkey} & {\scriptsize{}420.0000 } & {\scriptsize{}0.9833 } & {\scriptsize{}421.0000 } & {\scriptsize{}0.9738 } & {\scriptsize{}1.5146 }\tabularnewline
{\scriptsize{}Vietnam } & {\scriptsize{}495.0000 } & {\scriptsize{}0.9010 } & {\scriptsize{}494.0000 } & {\scriptsize{}0.8981 } & {\scriptsize{}1.0803 }\tabularnewline
\hline 
\end{tabular}{\scriptsize\par}
\par\end{center}

\begin{table}[H]
\caption{$\hat{\gamma}_{d}$ and $\hat{\gamma}_{d}^{\textsf{P}}$
along with their quality measures by country.\label{tab:Resultados-de-las estimaciones_y_medidas de calidad}}
\end{table}

Table \ref{tab:Resultados-de-las estimaciones_y_medidas de calidad}
shows the estimates $\hat{\gamma}_{d}$ of the mean ability by country
obtained in the Mathematics test along with their estimated coefficient of variation (CVE). 
This Table also displays the corresponding estimates according to the proposed estimator
$\hat{\gamma}_{d}^{P}$ together with the values of $EER_{d}=\frac{\sqrt{MSE}}{\hat{\gamma}_{d}^{P}}\times100\%$ as well as $Dif_{rel}=\frac{\left(\sigma_{d}^{2}-MSE\left(\hat{\gamma}_{d}^{P}\right)\right)}{\sigma_{d}^{2}}\times100\%$.
The latter is the relative difference between the measures of variability
of $\hat{\gamma}_{d}$ and $\hat{\gamma}_{d}^{P}$,
in order to compare the reduction in the variability of the estimation
of the mean ability in the Mathematics test using the proposed
estimator.

From Table \ref{tab:Resultados-de-las estimaciones_y_medidas de calidad},
we see that the $EER_{d}$, in all those countries that participated
in the test, are lower than the $CVE$ published by PISA. However,
it is not surprising that using the proposed estimator, the estimation
error decreased, as shown in the $Dif_{rel}\,\left(100\%\right)$ column,
even though they were already very small.

\section{Discussion}\label{sec_discussion}

This paper shows practical and methodological advantages of integrating small area estimation, item response theory, and multiple imputation. 
It is possible to show from a theoretical and empirical perspective that the proposed estimator for the ability mean is unbiased. 
In addition, in all of the scenarios where the proposed estimator is tested, it has a lower
average relative standard error than its competitors, considering simple random sampling. 
On the one hand, by varying the sample fractions and the domain fractions, the proposed estimator has lower mean relative standard errors compared to the other estimators used in the simulation. 
This behaviour is achieved under all the correlation scenarios.

Finally, when the percentage of missing data varies, the relative standard errors increases as the rate of missing data does. The proposed estimator always obtains the
lowest relative standard errors compared to the other estimators studied
in the simulation. Moreover, in the scenario with 30\% of missing
data and a small sampling fraction, the relative standard
errors did not exceed 5\% on average, implying that these estimates are quite low.

\bibliography{references_revision_juan.bib}
\bibliographystyle{apalike}

\appendix

\section{More on simulation study results}

\begin{table}[H]
\begin{centering}
{\scriptsize{}}%
\begin{tabular}{ccccccc}
\hline 
{\scriptsize{}$f_{d}$ (\%) } & {\scriptsize{}$f_{n}$(\%) } & {\scriptsize{}$\overline{EERP}$$\hat{\gamma}_{d}^{Dir}$(\%)} & {\scriptsize{}$\overline{EERP}$$\hat{\gamma}_{d}^{Cal}$(\%)} & {\scriptsize{}$\overline{EERP}$$\hat{\gamma}_{d}^{Comp}$ (\%)} & {\scriptsize{}$\overline{EERP}$$\hat{\gamma}_{d}^{P}$(\%)} & {\scriptsize{}$\overline{SBR}$$\hat{\gamma}_{d}^{P}$ (\%)}\tabularnewline
\hline 
{\scriptsize{}30\%} & {\scriptsize{}5\%} & {\scriptsize{}1.48} & {\scriptsize{}1.47} & {\scriptsize{}1.01} & {\scriptsize{}0.92} & {\scriptsize{}0.06}\tabularnewline
{\scriptsize{}30\%} & {\scriptsize{}10\%} & {\scriptsize{}1.17} & {\scriptsize{}1.15} & {\scriptsize{}0.94} & {\scriptsize{}0.88} & {\scriptsize{}0.05}\tabularnewline
{\scriptsize{}30\%} & {\scriptsize{}20\%} & {\scriptsize{}0.96} & {\scriptsize{}0.96} & {\scriptsize{}0.90} & {\scriptsize{}0.87} & {\scriptsize{}0.01}\tabularnewline
{\scriptsize{}50\%} & {\scriptsize{}5\%} & {\scriptsize{}1.84} & {\scriptsize{}1.84} & {\scriptsize{}1.11} & {\scriptsize{}0.96} & {\scriptsize{}0.09}\tabularnewline
{\scriptsize{}50\%} & {\scriptsize{}10\%} & {\scriptsize{}1.38} & {\scriptsize{}1.37} & {\scriptsize{}0.99} & {\scriptsize{}0.88} & {\scriptsize{}0.07}\tabularnewline
{\scriptsize{}50\%} & {\scriptsize{}20\%} & {\scriptsize{}1.09} & {\scriptsize{}1.09} & {\scriptsize{}0.92} & {\scriptsize{}0.86} & {\scriptsize{}0.04}\tabularnewline
{\scriptsize{}70\%} & {\scriptsize{}5\%} & {\scriptsize{}2.11} & {\scriptsize{}2.17} & {\scriptsize{}1.22} & {\scriptsize{}1.11} & {\scriptsize{}0.11}\tabularnewline
{\scriptsize{}70\%} & {\scriptsize{}10\%} & {\scriptsize{}1.58} & {\scriptsize{}1.57} & {\scriptsize{}1.04} & {\scriptsize{}0.92} & {\scriptsize{}0.08}\tabularnewline
{\scriptsize{}70\%} & {\scriptsize{}20\%} & {\scriptsize{}1.22} & {\scriptsize{}1.21} & {\scriptsize{}0.95} & {\scriptsize{}0.87} & {\scriptsize{}0.05}\tabularnewline
\hline 
\end{tabular}{\scriptsize\par}
\par\end{centering}
\caption{Missing 10\% and medium correlation\label{Tabla 3_media}}
\end{table}

\begin{table}[H]
\begin{centering}
{\scriptsize{}}%
\begin{tabular}{ccccccc}
\hline 
{\scriptsize{}$f_{d}$ (\%) } & {\scriptsize{}$f_{n}$(\%) } & {\scriptsize{}$\overline{EERP}$$\hat{\gamma}_{d}^{Dir}$(\%)} & {\scriptsize{}$\overline{EERP}$$\hat{\gamma}_{d}^{Cal}$(\%)} & {\scriptsize{}$\overline{EERP}$$\hat{\gamma}_{d}^{Comp}$ (\%)} & {\scriptsize{}$\overline{EERP}$$\hat{\gamma}_{d}^{P}$(\%)} & {\scriptsize{}$\overline{SBR}$$\hat{\gamma}_{d}^{P}$ (\%)}\tabularnewline
\hline 
{\scriptsize{}30\%} & {\scriptsize{}5\%} & {\scriptsize{}1.49} & {\scriptsize{}2.84} & {\scriptsize{}1.33} & {\scriptsize{}1.22} & {\scriptsize{}-0.06}\tabularnewline
{\scriptsize{}30\%} & {\scriptsize{}10\%} & {\scriptsize{}1.19} & {\scriptsize{}1.93} & {\scriptsize{}1.10} & {\scriptsize{}1.08} & {\scriptsize{}-0.03}\tabularnewline
{\scriptsize{}30\%} & {\scriptsize{}20\%} & {\scriptsize{}1.00} & {\scriptsize{}1.24} & {\scriptsize{}0.97} & {\scriptsize{}1.01} & {\scriptsize{}0.01}\tabularnewline
{\scriptsize{}50\%} & {\scriptsize{}5\%} & {\scriptsize{}1.84} & {\scriptsize{}3.85} & {\scriptsize{}1.61} & {\scriptsize{}1.45} & {\scriptsize{}-0.14}\tabularnewline
{\scriptsize{}50\%} & {\scriptsize{}10\%} & {\scriptsize{}1.40} & {\scriptsize{}2.60} & {\scriptsize{}1.26} & {\scriptsize{}1.15} & {\scriptsize{}-0.04}\tabularnewline
{\scriptsize{}50\%} & {\scriptsize{}20\%} & {\scriptsize{}1.13} & {\scriptsize{}1.72} & {\scriptsize{}1.06} & {\scriptsize{}1.01} & {\scriptsize{}0.01}\tabularnewline
{\scriptsize{}70\%} & {\scriptsize{}5\%} & {\scriptsize{}2.12} & {\scriptsize{}4.71} & {\scriptsize{}1.95} & {\scriptsize{}1.86} & {\scriptsize{}-0.19}\tabularnewline
{\scriptsize{}70\%} & {\scriptsize{}10\%} & {\scriptsize{}1.58} & {\scriptsize{}3.12} & {\scriptsize{}1.40} & {\scriptsize{}1.23} & {\scriptsize{}-0.11}\tabularnewline
{\scriptsize{}70\%} & {\scriptsize{}20\%} & {\scriptsize{}1.24} & {\scriptsize{}2.11} & {\scriptsize{}1.14} & {\scriptsize{}1.03} & {\scriptsize{}-0.02}\tabularnewline
\hline 
\end{tabular}{\scriptsize\par}
\par\end{centering}
\caption{Missing 10\% and low correlation \label{Tabla 4_media}}
\end{table}

\begin{table}[H]
\begin{centering}
{\scriptsize{}}%
\begin{tabular}{ccccccc}
\hline 
{\scriptsize{}$f_{d}$ (\%) } & {\scriptsize{}$f_{n}$(\%) } & {\scriptsize{}$\overline{EERP}$$\hat{\gamma}_{d}^{Dir}$(\%)} & {\scriptsize{}$\overline{EERP}$$\hat{\gamma}_{d}^{Cal}$(\%)} & {\scriptsize{}$\overline{EERP}$$\hat{\gamma}_{d}^{Comp}$ (\%)} & {\scriptsize{}$\overline{EERP}$$\hat{\gamma}_{d}^{P}$(\%)} & {\scriptsize{}$\overline{SBR}$$\hat{\gamma}_{d}^{P}$ (\%)}\tabularnewline
\hline 
{\scriptsize{}30\%} & {\scriptsize{}5\%} & {\scriptsize{}1.57} & {\scriptsize{}1.25} & {\scriptsize{}1.09} & {\scriptsize{}0.92} & {\scriptsize{}0.07}\tabularnewline
{\scriptsize{}30\%} & {\scriptsize{}10\%} & {\scriptsize{}1.27} & {\scriptsize{}1.12} & {\scriptsize{}1.05} & {\scriptsize{}0.91} & {\scriptsize{}0.02}\tabularnewline
{\scriptsize{}30\%} & {\scriptsize{}20\%} & {\scriptsize{}1.10} & {\scriptsize{}1.06} & {\scriptsize{}1.04} & {\scriptsize{}0.91} & {\scriptsize{}-0.01}\tabularnewline
{\scriptsize{}50\%} & {\scriptsize{}5\%} & {\scriptsize{}1.90} & {\scriptsize{}1.42} & {\scriptsize{}1.14} & {\scriptsize{}0.94} & {\scriptsize{}0.03}\tabularnewline
{\scriptsize{}50\%} & {\scriptsize{}10\%} & {\scriptsize{}1.48} & {\scriptsize{}1.20} & {\scriptsize{}1.08} & {\scriptsize{}0.93} & {\scriptsize{}0.06}\tabularnewline
{\scriptsize{}50\%} & {\scriptsize{}20\%} & {\scriptsize{}1.22} & {\scriptsize{}1.10} & {\scriptsize{}1.05} & {\scriptsize{}0.92} & {\scriptsize{}0.01}\tabularnewline
{\scriptsize{}70\%} & {\scriptsize{}5\%} & {\scriptsize{}2.18} & {\scriptsize{}1.59} & {\scriptsize{}1.20} & {\scriptsize{}1.02} & {\scriptsize{}0.24}\tabularnewline
{\scriptsize{}70\%} & {\scriptsize{}10\%} & {\scriptsize{}1.67} & {\scriptsize{}1.29} & {\scriptsize{}1.10} & {\scriptsize{}0.94} & {\scriptsize{}0.1}\tabularnewline
{\scriptsize{}70\%} & {\scriptsize{}20\%} & {\scriptsize{}1.33} & {\scriptsize{}1.14} & {\scriptsize{}1.06} & {\scriptsize{}0.95} & {\scriptsize{}0.03}\tabularnewline
\hline 
\end{tabular}{\scriptsize\par}
\par\end{centering}
\caption{Missing 20\% and high correlation. \label{Tabla 5_media}}
\end{table}

\begin{table}[H]
\begin{centering}
{\scriptsize{}}%
\begin{tabular}{ccccccc}
\hline 
{\scriptsize{}$f_{d}$ (\%) } & {\scriptsize{}$f_{n}$(\%) } & {\scriptsize{}$\overline{EERP}$$\hat{\gamma}_{d}^{Dir}$(\%)} & {\scriptsize{}$\overline{EERP}$$\hat{\gamma}_{d}^{Cal}$(\%)} & {\scriptsize{}$\overline{EERP}$$\hat{\gamma}_{d}^{Comp}$ (\%)} & {\scriptsize{}$\overline{EERP}$$\hat{\gamma}_{d}^{P}$(\%)} & {\scriptsize{}$\overline{SBR}$$\hat{\gamma}_{d}^{P}$ (\%)}\tabularnewline
\hline 
{\scriptsize{}30\%} & {\scriptsize{}5\%} & {\scriptsize{}1.53} & {\scriptsize{}1.52} & {\scriptsize{}1.09} & {\scriptsize{}0.98} & {\scriptsize{}0.07}\tabularnewline
{\scriptsize{}30\%} & {\scriptsize{}10\%} & {\scriptsize{}1.22} & {\scriptsize{}1.22} & {\scriptsize{}1.02} & {\scriptsize{}0.95} & {\scriptsize{}0.04}\tabularnewline
{\scriptsize{}30\%} & {\scriptsize{}20\%} & {\scriptsize{}1.04} & {\scriptsize{}1.04} & {\scriptsize{}0.98} & {\scriptsize{}0.94} & {\scriptsize{}0.00}\tabularnewline
{\scriptsize{}50\%} & {\scriptsize{}5\%} & {\scriptsize{}1.85} & {\scriptsize{}1.89} & {\scriptsize{}1.18} & {\scriptsize{}1.04} & {\scriptsize{}0.09}\tabularnewline
{\scriptsize{}50\%} & {\scriptsize{}10\%} & {\scriptsize{}1.44} & {\scriptsize{}1.43} & {\scriptsize{}1.08} & {\scriptsize{}0.97} & {\scriptsize{}0.07}\tabularnewline
{\scriptsize{}50\%} & {\scriptsize{}20\%} & {\scriptsize{}1.17} & {\scriptsize{}1.16} & {\scriptsize{}1.01} & {\scriptsize{}0.94} & {\scriptsize{}0.03}\tabularnewline
{\scriptsize{}70\%} & {\scriptsize{}5\%} & {\scriptsize{}2.13} & {\scriptsize{}2.24} & {\scriptsize{}1.29} & {\scriptsize{}1.19} & {\scriptsize{}0.09}\tabularnewline
{\scriptsize{}70\%} & {\scriptsize{}10\%} & {\scriptsize{}1.62} & {\scriptsize{}1.61} & {\scriptsize{}1.11} & {\scriptsize{}0.98} & {\scriptsize{}0.09}\tabularnewline
{\scriptsize{}70\%} & {\scriptsize{}20\%} & {\scriptsize{}1.28} & {\scriptsize{}1.27} & {\scriptsize{}1.03} & {\scriptsize{}0.95} & {\scriptsize{}0.05}\tabularnewline
\hline 
\end{tabular}{\scriptsize\par}
\par\end{centering}
\caption{Missing 20\% and correlation medium \label{Tabla 6_media}}
\end{table}
\begin{table}[H]
\begin{centering}
{\scriptsize{}}%
\begin{tabular}{ccccccc}
\hline 
{\scriptsize{}$f_{d}$ (\%) } & {\scriptsize{}$f_{n}$(\%) } & {\scriptsize{}$\overline{EERP}$$\hat{\gamma}_{d}^{Dir}$(\%)} & {\scriptsize{}$\overline{EERP}$$\hat{\gamma}_{d}^{Cal}$(\%)} & {\scriptsize{}$\overline{EERP}$$\hat{\gamma}_{d}^{Comp}$ (\%)} & {\scriptsize{}$\overline{EERP}$$\hat{\gamma}_{d}^{P}$(\%)} & {\scriptsize{}$\overline{SBR}$$\hat{\gamma}_{d}^{P}$ (\%)}\tabularnewline
\hline 
{\scriptsize{}30\%} & {\scriptsize{}5\%} & {\scriptsize{}1.52} & {\scriptsize{}2.83} & {\scriptsize{}1.34} & {\scriptsize{}1.16} & {\scriptsize{}-0.07}\tabularnewline
{\scriptsize{}30\%} & {\scriptsize{}10\%} & {\scriptsize{}1.20} & {\scriptsize{}1.93} & {\scriptsize{}1.12} & {\scriptsize{}1.02} & {\scriptsize{}0.01}\tabularnewline
{\scriptsize{}30\%} & {\scriptsize{}20\%} & {\scriptsize{}1.02} & {\scriptsize{}1.26} & {\scriptsize{}1.00} & {\scriptsize{}0.96} & {\scriptsize{}0.02}\tabularnewline
{\scriptsize{}50\%} & {\scriptsize{}5\%} & {\scriptsize{}1.83} & {\scriptsize{}3.83} & {\scriptsize{}1.61} & {\scriptsize{}1.41} & {\scriptsize{}-0.07}\tabularnewline
{\scriptsize{}50\%} & {\scriptsize{}10\%} & {\scriptsize{}1.41} & {\scriptsize{}2.57} & {\scriptsize{}1.27} & {\scriptsize{}1.08} & {\scriptsize{}-0.04}\tabularnewline
{\scriptsize{}50\%} & {\scriptsize{}20\%} & {\scriptsize{}1.15} & {\scriptsize{}1.73} & {\scriptsize{}1.08} & {\scriptsize{}0.96} & {\scriptsize{}0.01}\tabularnewline
{\scriptsize{}70\%} & {\scriptsize{}5\%} & {\scriptsize{}2.12} & {\scriptsize{}4.70} & {\scriptsize{}1.92} & {\scriptsize{}1.86} & {\scriptsize{}-0.16}\tabularnewline
{\scriptsize{}70\%} & {\scriptsize{}10\%} & {\scriptsize{}1.60} & {\scriptsize{}3.11} & {\scriptsize{}1.41} & {\scriptsize{}1.19} & {\scriptsize{}-0.08}\tabularnewline
{\scriptsize{}70\%} & {\scriptsize{}20\%} & {\scriptsize{}1.26} & {\scriptsize{}2.11} & {\scriptsize{}1.16} & {\scriptsize{}0.99} & {\scriptsize{}-0.01}\tabularnewline
\hline 
\end{tabular}{\scriptsize\par}
\par\end{centering}
\caption{Missing 20\% and low correlation \label{Tabla 7_media}}
\end{table}
\begin{table}[H]
\begin{centering}
{\scriptsize{}}%
\begin{tabular}{ccccccc}
\hline 
{\scriptsize{}$f_{d}$ (\%) } & {\scriptsize{}$f_{n}$(\%) } & {\scriptsize{}$\overline{EERP}$$\hat{\gamma}_{d}^{Dir}$(\%)} & {\scriptsize{}$\overline{EERP}$$\hat{\gamma}_{d}^{Cal}$(\%)} & {\scriptsize{}$\overline{EERP}$$\hat{\gamma}_{d}^{Comp}$ (\%)} & {\scriptsize{}$\overline{EERP}$$\hat{\gamma}_{d}^{P}$(\%)} & {\scriptsize{}$\overline{SBR}$$\hat{\gamma}_{d}^{P}$ (\%)}\tabularnewline
\hline 
{\scriptsize{}30\%} & {\scriptsize{}5\%} & {\scriptsize{}1.57} & {\scriptsize{}1.26} & {\scriptsize{}1.11} & {\scriptsize{}0.90} & {\scriptsize{}0.1}\tabularnewline
{\scriptsize{}30\%} & {\scriptsize{}10\%} & {\scriptsize{}1.29} & {\scriptsize{}1.13} & {\scriptsize{}1.07} & {\scriptsize{}0.89} & {\scriptsize{}0.04}\tabularnewline
{\scriptsize{}30\%} & {\scriptsize{}20\%} & {\scriptsize{}1.11} & {\scriptsize{}1.07} & {\scriptsize{}1.05} & {\scriptsize{}0.90} & {\scriptsize{}0.01}\tabularnewline
{\scriptsize{}50\%} & {\scriptsize{}5\%} & {\scriptsize{}1.89} & {\scriptsize{}1.41} & {\scriptsize{}1.16} & {\scriptsize{}0.92} & {\scriptsize{}0.17}\tabularnewline
{\scriptsize{}50\%} & {\scriptsize{}10\%} & {\scriptsize{}1.48} & {\scriptsize{}1.21} & {\scriptsize{}1.10} & {\scriptsize{}0.90} & {\scriptsize{}0.07}\tabularnewline
{\scriptsize{}50\%} & {\scriptsize{}20\%} & {\scriptsize{}1.23} & {\scriptsize{}1.11} & {\scriptsize{}1.07} & {\scriptsize{}0.92} & {\scriptsize{}0.03}\tabularnewline
{\scriptsize{}70\%} & {\scriptsize{}5\%} & {\scriptsize{}2.16} & {\scriptsize{}1.58} & {\scriptsize{}1.21} & {\scriptsize{}1.01} & {\scriptsize{}0.25}\tabularnewline
{\scriptsize{}70\%} & {\scriptsize{}10\%} & {\scriptsize{}1.67} & {\scriptsize{}1.29} & {\scriptsize{}1.12} & {\scriptsize{}0.92} & {\scriptsize{}0.11}\tabularnewline
{\scriptsize{}70\%} & {\scriptsize{}20\%} & {\scriptsize{}1.34} & {\scriptsize{}1.15} & {\scriptsize{}1.08} & {\scriptsize{}0.93} & {\scriptsize{}0.05}\tabularnewline
\hline 
\end{tabular}{\scriptsize\par}
\par\end{centering}
\caption{Missing 30\% and High correlation \label{tabla 8_media}}
\end{table}
\begin{table}[H]
\begin{centering}
{\scriptsize{}}%
\begin{tabular}{ccccccc}
\hline 
{\scriptsize{}$f_{d}$ (\%) } & {\scriptsize{}$f_{n}$(\%) } & {\scriptsize{}$\overline{EERP}$$\hat{\gamma}_{d}^{Dir}$(\%)} & {\scriptsize{}$\overline{EERP}$$\hat{\gamma}_{d}^{Cal}$(\%)} & {\scriptsize{}$\overline{EERP}$$\hat{\gamma}_{d}^{Comp}$ (\%)} & {\scriptsize{}$\overline{EERP}$$\hat{\gamma}_{d}^{P}$(\%)} & {\scriptsize{}$\overline{SBR}$$\hat{\gamma}_{d}^{P}$ (\%)}\tabularnewline
\hline 
{\scriptsize{}30\%} & {\scriptsize{}5\%} & {\scriptsize{}1.64} & {\scriptsize{}1.65} & {\scriptsize{}1.24} & {\scriptsize{}1.04} & {\scriptsize{}0.09}\tabularnewline
{\scriptsize{}30\%} & {\scriptsize{}10\%} & {\scriptsize{}1.36} & {\scriptsize{}1.36} & {\scriptsize{}1.18} & {\scriptsize{}1.04} & {\scriptsize{}0.04}\tabularnewline
{\scriptsize{}30\%} & {\scriptsize{}20\%} & {\scriptsize{}1.20} & {\scriptsize{}1.20} & {\scriptsize{}1.15} & {\scriptsize{}1.03} & {\scriptsize{}0.02}\tabularnewline
{\scriptsize{}50\%} & {\scriptsize{}5\%} & {\scriptsize{}1.94} & {\scriptsize{}2.00} & {\scriptsize{}1.33} & {\scriptsize{}1.11} & {\scriptsize{}0.12}\tabularnewline
{\scriptsize{}50\%} & {\scriptsize{}10\%} & {\scriptsize{}1.54} & {\scriptsize{}1.57} & {\scriptsize{}1.22} & {\scriptsize{}1.04} & {\scriptsize{}0.07}\tabularnewline
{\scriptsize{}50\%} & {\scriptsize{}20\%} & {\scriptsize{}1.31} & {\scriptsize{}1.31} & {\scriptsize{}1.17} & {\scriptsize{}1.03} & {\scriptsize{}0.04}\tabularnewline
{\scriptsize{}70\%} & {\scriptsize{}5\%} & {\scriptsize{}2.22} & {\scriptsize{}2.34} & {\scriptsize{}1.42} & {\scriptsize{}1.25} & {\scriptsize{}0.13}\tabularnewline
{\scriptsize{}70\%} & {\scriptsize{}10\%} & {\scriptsize{}1.72} & {\scriptsize{}1.74} & {\scriptsize{}1.26} & {\scriptsize{}1.07} & {\scriptsize{}0.09}\tabularnewline
{\scriptsize{}70\%} & {\scriptsize{}20\%} & {\scriptsize{}1.41} & {\scriptsize{}1.42} & {\scriptsize{}1.19} & {\scriptsize{}1.05} & {\scriptsize{}0.06}\tabularnewline
\hline 
\end{tabular}{\scriptsize\par}
\par\end{centering}
\caption{Missing 30\% and correlation medium \label{Tabla 9_media}}
\end{table}
\begin{table}[H]
\begin{centering}
{\scriptsize{}}%
\begin{tabular}{ccccccc}
\hline 
{\scriptsize{}$f_{d}$ (\%) } & {\scriptsize{}$f_{n}$(\%) } & {\scriptsize{}$\overline{EERP}$$\hat{\gamma}_{d}^{Dir}$(\%)} & {\scriptsize{}$\overline{EERP}$$\hat{\gamma}_{d}^{Cal}$(\%)} & {\scriptsize{}$\overline{EERP}$$\hat{\gamma}_{d}^{Comp}$ (\%)} & {\scriptsize{}$\overline{EERP}$$\hat{\gamma}_{d}^{P}$(\%)} & {\scriptsize{}$\overline{SBR}$$\hat{\gamma}_{d}^{P}$ (\%)}\tabularnewline
\hline 
{\scriptsize{}30\%} & {\scriptsize{}5\%} & {\scriptsize{}1.60} & {\scriptsize{}2.95} & {\scriptsize{}1.45} & {\scriptsize{}1.24} & {\scriptsize{}-0.07}\tabularnewline
{\scriptsize{}30\%} & {\scriptsize{}10\%} & {\scriptsize{}1.32} & {\scriptsize{}2.03} & {\scriptsize{}1.24} & {\scriptsize{}1.08} & {\scriptsize{}-0.01}\tabularnewline
{\scriptsize{}30\%} & {\scriptsize{}20\%} & {\scriptsize{}1.16} & {\scriptsize{}1.39} & {\scriptsize{}1.14} & {\scriptsize{}1.03} & {\scriptsize{}0.02}\tabularnewline
{\scriptsize{}50\%} & {\scriptsize{}5\%} & {\scriptsize{}1.91} & {\scriptsize{}3.93} & {\scriptsize{}1.70} & {\scriptsize{}1.46} & {\scriptsize{}-0.14}\tabularnewline
{\scriptsize{}50\%} & {\scriptsize{}10\%} & {\scriptsize{}1.51} & {\scriptsize{}2.67} & {\scriptsize{}1.38} & {\scriptsize{}1.14} & {\scriptsize{}-0.05}\tabularnewline
{\scriptsize{}50\%} & {\scriptsize{}20\%} & {\scriptsize{}1.26} & {\scriptsize{}1.84} & {\scriptsize{}1.21} & {\scriptsize{}1.04} & {\scriptsize{}0.00}\tabularnewline
{\scriptsize{}70\%} & {\scriptsize{}5\%} & {\scriptsize{}2.18} & {\scriptsize{}4.83} & {\scriptsize{}2.00} & {\scriptsize{}1.94} & {\scriptsize{}-0.14}\tabularnewline
{\scriptsize{}70\%} & {\scriptsize{}10\%} & {\scriptsize{}1.68} & {\scriptsize{}3.20} & {\scriptsize{}1.51} & {\scriptsize{}1.26} & {\scriptsize{}-0.08}\tabularnewline
{\scriptsize{}70\%} & {\scriptsize{}20\%} & {\scriptsize{}1.37} & {\scriptsize{}2.20} & {\scriptsize{}1.28} & {\scriptsize{}1.07} & {\scriptsize{}-0.01}\tabularnewline
\hline 
\end{tabular}{\scriptsize\par}
\par\end{centering}
\caption{Missing 30\% and low correlation \label{Tabla 10_media}}
\end{table}

\section{Notation}

Matrices and vectors with entries consisting of subscripted variables are denoted by a boldfaced version of the letter for that variable. For example, $\boldsymbol{x} = (x_1,\ldots, x_n)$ denotes an $n\times1$ column vector with entries $x_1,\ldots, x_n$. We use $\boldsymbol{0}$ and $\boldsymbol{1}$ to denote the column vector with all entries equal to 0 and 1, respectively, and $\mathbf{I}$ to denote the identity matrix. A subindex in this context refers to the corresponding dimension; for instance, $\mathbf{I}_n$ denotes the $n\times n$ identity matrix. The transpose of a vector $\boldsymbol{x}$ is denoted by $\boldsymbol{x}^{\textsf{T}}$; analogously for matrices. Moreover, if $\mathbf{X}$ is a square matrix, we use $\text{tr}(\mathbf{X})$ to denote its trace and $\mathbf{X}^{-1}$ to denote its inverse. The norm of $\boldsymbol{x}$, given by $\sqrt{\boldsymbol{x}^{\textsf{T}}\boldsymbol{x}}$, is denoted by $\|\,\boldsymbol{x}\|\,$.

\end{document}